%% file: main.tex
\documentclass[10pt,conference]{IEEEtran} 
\usepackage[colorlinks=true, linkcolor=blue, 
            citecolor=blue, urlcolor=blue,
            bookmarks=false,        %
            bookmarksnumbered=false,%
]{hyperref}
\usepackage[noadjust]{cite}
\usepackage{multirow}
\usepackage{subfigure}
\usepackage{rotating}
\usepackage{array}
\usepackage[disable]{todonotes}
\usepackage{xspace}
\usepackage{etoolbox}
\usepackage{amsmath}
\usepackage{soul}
\usepackage{cleveref}
\usepackage{pdfpages}
\usepackage{flushend}
\usepackage{footnote}
\def\BibTeX{{\rm B\kern-.05em{\sc i\kern-.025em b}\kern-.08em
    T\kern-.1667em\lower.7ex\hbox{E}\kern-.125emX}}

\makeatletter
\patchcmd{\@makecaption}
  {\scshape}
  {}
  {}
  {}
\makeatother

\newcommand{\mytt}[1]{{{\tt #1}}}
\newcommand{\projname}{{\sc Holmes}\xspace}

\newcommand{\rtodo}[1]{\todo[inline]{Rigel: #1}}

\newcolumntype{M}[1]{>{\centering\arraybackslash}p{#1}}

\usepackage{enumitem}
\setlist[itemize,1]{leftmargin=1.1\parindent, itemsep=0.7ex, topsep=0.3ex, %
}
\setlist[enumerate,1]{leftmargin=2\parindent, itemsep=0ex, topsep=0.5ex, %
}

\makeatletter
\renewcommand*{\thetable}{\arabic{table}}
\renewcommand*{\thefigure}{\arabic{figure}}
\let\c@table\c@figure
\makeatother

\begin{document}

\newcommand{\titleB}{HOLMES: Real-time APT Detection through Correlation of Suspicious Information Flows}
\date{}

\twocolumn[{%
 \centering
 \huge \textbf{\titleB}\\[1em]
 \large Sadegh M. Milajerdi\IEEEauthorrefmark{1},
        Rigel Gjomemo\IEEEauthorrefmark{1},
        Birhanu Eshete\IEEEauthorrefmark{2}$^,$\footnotemark,
        R. Sekar\IEEEauthorrefmark{3},
        V.N. Venkatakrishnan\IEEEauthorrefmark{1}\\[1em]
 \normalsize
 \begin{tabular}{*{3}{>{\centering}p{.3\textwidth}}}
  \IEEEauthorrefmark{1}University of Illinois at Chicago & \IEEEauthorrefmark{2}University of Michigan-Dearborn & \IEEEauthorrefmark{3}Stony Brook University \tabularnewline
  \{smomen2,rgjome1,venkat\}@uic.edu & birhanu@umich.edu &  sekar@cs.stonybrook.edu
 \end{tabular}\\[2em] %
}]
\thispagestyle{empty}
\footnotetext[1]{The third author performed this work as a postdoctoral associate at the University of Illinois at Chicago.}
\begin{abstract}
\boldmath
In this paper, we present \projname, a system that implements a new approach to the detection of Advanced and Persistent Threats (APTs). \projname is inspired by several case studies of real-world APTs that  highlight some  common goals  of APT actors.   In a nutshell, \projname aims to produce a   detection signal that indicates the presence of a coordinated set of activities that are part of an APT  campaign.    One of the main challenges addressed by our approach involves  developing a suite of  techniques that make the detection signal robust and reliable. At a high-level, the techniques we develop effectively leverage the {\em correlation between  suspicious information flows} that arise during an attacker campaign.  In addition to its detection capability, \projname is also able to generate a high-level graph that summarizes the attacker's actions in real-time. This graph can be used by an analyst for an effective cyber response. 
An  evaluation of our approach against some real-world APTs indicates that \projname can detect APT campaigns with high precision and low false alarm rate.  The compact high-level graphs produced by \projname effectively summarizes an ongoing attack campaign and can assist real-time cyber-response operations.

\end{abstract}

\input{intro}
\input{background}
\input{approach}

\input{design}
\input{implementation}
\input{eval}
\input{relwork}

\input{conclusion}
\section*{Acknowledgments}

We thank Guofei Gu for the helpful review comments and suggestions to the manuscript. This work was primarily supported by DARPA (under AFOSR contract
FA8650-15-C-7561) and in part by SPAWAR (N6600118C4035), NSF (CNS-1319137,
CNS-1514472, and DGE-1069311), and ONR (N00014-15-1-2378, and N00014-17-1-2891).
The views, opinions, and/or findings expressed are those of the authors and
should not be interpreted as representing the official views or policies of the
Department of Defense, National Science Foundation or the U.S. Government.

\bibliographystyle{plain}
\bibliography{bib}

\input{appendix}

\end{document}

%% file: intro.tex
\section{Introduction} \label{sec:intro}

 In one of the first ever detailed reports on  Advanced and Persistent Threats (entitled APT1~\cite{mandiant-report}), the security firm Mandiant disclosed the goals and activities of a global APT actor. The activities included stealing of hundreds of terabytes of sensitive data (including business plans, technology blueprints, and test results) from at least 141 organizations across a diverse set of industries. They estimated the average duration of persistence of malware in the targeted organizations to be 365 days. Since then, there has  been a growing list of documented APTs involving powerful actors, including nation-state actors, on the global scene.

\begin{figure}[ht!]
  \begin{center}
    \includegraphics[width=\columnwidth]{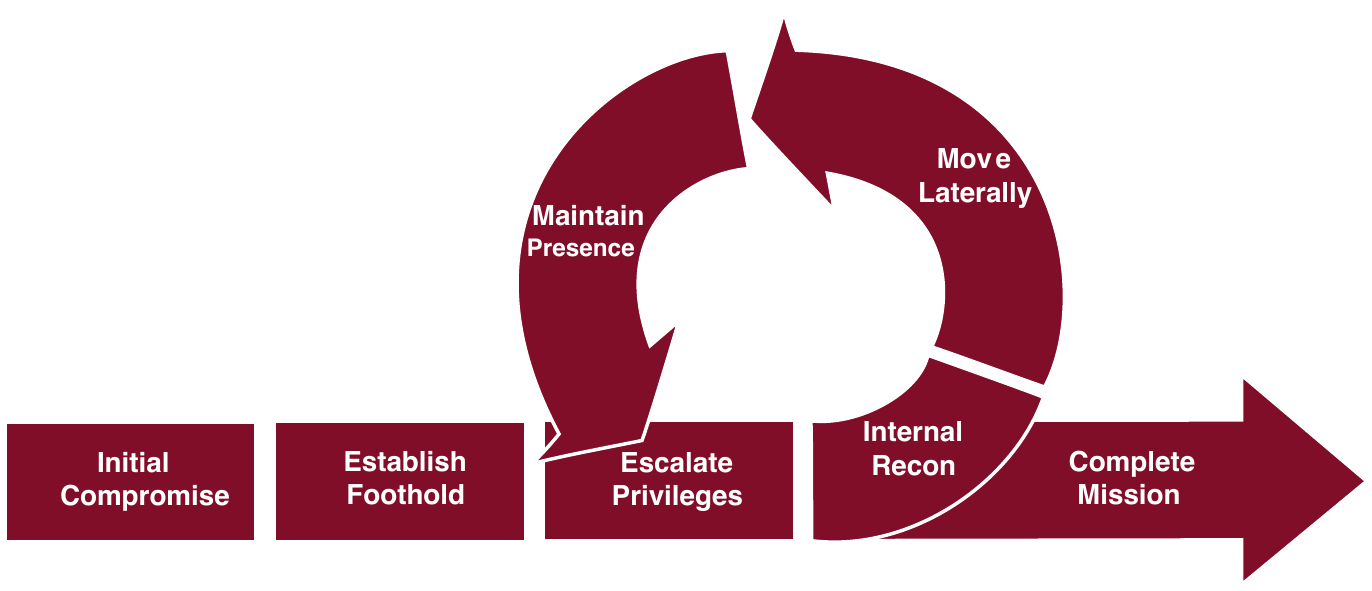}
    \vspace*{-2em}
     \caption{APT Lifecycle.}
     \vspace*{-2em}
    \label{fig:apt-lifecycle}
  \end{center}
  
\end{figure}

Understanding the motivations and operations of the APT actors plays a vital role in the challenge of  addressing these threats. To further this understanding, the Mandiant report also offered an APT lifecycle model (Fig.~\ref{fig:apt-lifecycle}), also known as the {\em kill-chain}, that allows one to gain perspective on how the APT steps collectively achieve their actors' goals.  A typical APT attack consists of a successful penetration (e.g., a drive-by-download or a spear-phishing attack), reconnaissance, command and control (C\&C) communication (sometimes using Remote Access Trojans (RATs)), privilege escalation (by exploiting vulnerabilities), lateral movement through  the network, exfiltration of confidential information, and so on. In short, the  kill-chain provides a reference  to understand and map the motivations, targets, and actions of APT actors.

APTs have grown in sophistication since the publication of the first Mandiant report. The details of various exploits used have  varied over the years, but the high-level steps have remained mostly the same. While surveying about 300 APT reports~\cite{apt-reports}, we observed that 
\begin{itemize}
\item the goal of an APT actor is either to obtain and exfiltrate highly confidential information, e.g., source code of specific proprietary technology; or to damage the victim by compromising  high-integrity resources, e.g., PLCs compromised in the Stuxnet worm, and 
\item this goal is accomplished primarily through steps that conform to the  kill-chain shown in Fig.~\ref{fig:apt-lifecycle}. 
\end{itemize}

Existing IDS/IPS systems in an enterprise may detect and produce alerts for suspicious events on a host. However, combining these low-level alerts to derive a high-level picture of an ongoing APT campaign remains a major challenge. \rtodo{this paragraph feels a bit standalone. Incorporate into State of the Art below?}

{\em State of the art.} Today, alert correlation is typically performed using Security Information and Event Management (SIEM) systems such as Splunk~\cite{splunk}, LogRhythm \cite{logrhythm} and IBM QRadar \cite{QRadar}. These systems collect log events and alerts from multiple sources and correlate them. Such correlation often makes use of readily available indicators,  such as timestamps for instance. These correlation methods are useful, but they often lack (a) an understanding of the complex relationships that exist between alerts and actual intrusion instances and (b) the precision needed to piece together attack steps that take place on different hosts over long periods of time (weeks, or in some cases, months). %

\noindent{\bf Problem Statement. } 
{\em The main problem tackled in this paper is to detect  an ongoing APT campaign (that consists of many disparate steps across many hosts  over a long period of time) in real-time and provide a high-level explanation of the attack scenario to an analyst, based on host logs and IPS alerts  from the enterprise. }

There are three main aspects to this problem, and they are as follows:

\begin{itemize}
\item Alert generation: Starting from low-level event traces from hosts, we must generate alerts in an efficient manner. How do we generate alerts that attempt to factor any  significant steps the attacker might be taking? Additionally, care must be taken to ensure that we do not generate a large volume of noisy alerts. 

\item Alert correlation: The challenge here is to combine these alerts from multiple activities of the attacker into a reliable signal that indicates the presence of an ongoing APT campaign.

\item Attack scenario  presentation: Indicators of an ongoing APT campaign needs to be communicated to a human being (a cyber-analyst). To be effective, this communication must be intuitive and needs to summarize the attack at a high level such that the analyst quickly realizes the scope and magnitude of the campaign. 
\end{itemize}

\noindent {\bf Approach and Contributions. }
We present a system called \projname in this paper that addresses all the above aspects. 
\projname begins with host audit data (e.g., Linux {\tt auditd} or Windows ETW data) and produces a detection signal that maps out the stages of an ongoing APT campaign. At a high level,  \projname makes {\em novel use of the APT kill-chain} as the pivotal reference in addressing the technical challenges involved in the above three aspects of APT detection. We describe our key ideas  and their significance below, with a detailed technical description appearing in Section~\ref{sec:approach}. 

First, \projname aims to map the activities found in host logs as well as any alerts found in the enterprise directly to the kill chain.  This design choice allows \projname to generate alerts that are semantically close to the activity steps (``Tactics, Techniques and Procedures'' (TTPs)) of APT actors. By doing so, \projname elevates the alert generation process to work at the level of the steps of an attack campaign, than about how they manifest in low-level audit logs. Thus, we solve an important challenge in generating alerts of significance. In our experiments, we have found that a five-day collection of audit logs contains around 3M low-level events, while \projname only extracts 86 suspicious activity steps from them.

A second important idea in \projname is to {\em use the information flow between
  low-level entities (files, processes, etc.) in the system as the basis for alert
  correlation}. To see this, note that the internal reconnaissance step in the
kill-chain depends on a successful initial compromise and establishment of a
foothold. In particular, the reconnaissance step is typically launched using the
command and control agent (process) installed by the attacker during foothold
establishment, thus exhibiting a flow between the processes involved in the two
phases. Moreover, reconnaissance often involves running malware (files)
downloaded during the foothold establishment phase, illustrating a
file-to-process flow. Similarly, a successful lateral movement phase, as well as
the exfiltration phase, uses data gathered by the reconnaissance phase. Thus, by
detecting low-level events associated with APT steps and linking them using
information flow, it is possible to construct the emerging kill-chain used by an
APT actor.

A third main contribution in \projname is the development of a high-level
scenario graph (HSG). The nodes of the HSG correspond to TTPs, and the edges
represent information flows between entities involved in the TTPs. The HSG
provides the basis for detecting APTs with high confidence. For this purpose, we
develop several new ideas. First is the concept of an {\em ancestral cover} in
an HSG. We show how this concept can help to assess the strength of dependencies
between HSG nodes. Weak dependencies can then be pruned away to eliminate many
false alarms. Second, we develop {\em noise reduction} techniques that further
de-emphasize dependencies that are known to be associated with benign
activities. Third, we develop ranking and prioritization techniques to prune
away most nodes and edges unrelated to the APT campaign. These steps are
described in detail in Sections~\ref{subsec:dep-expl}, \ref{subsec:noise}, and
\ref{subsec:detection}. Using these techniques, we  demonstrate that \projname is able to make  a clear distinction between attack and benign scenarios. 

Finally, the HSG provides a very compact, visual summary of the campaign at any
moment, thus making an important contribution for attack comprehension. For
instance, starting from a dataset of 10M audit records, we are
able to summarize a high-level attack campaign using a graph of just 16 nodes. A cyber-analyst can use the presented HSG to  quickly infer the big picture of the attack (scope and magnitude) with relative ease.  
\begin{figure*}[ht!]
  \begin{center}
    \hspace*{-0.5em}
    \includegraphics[width=2.1\columnwidth]{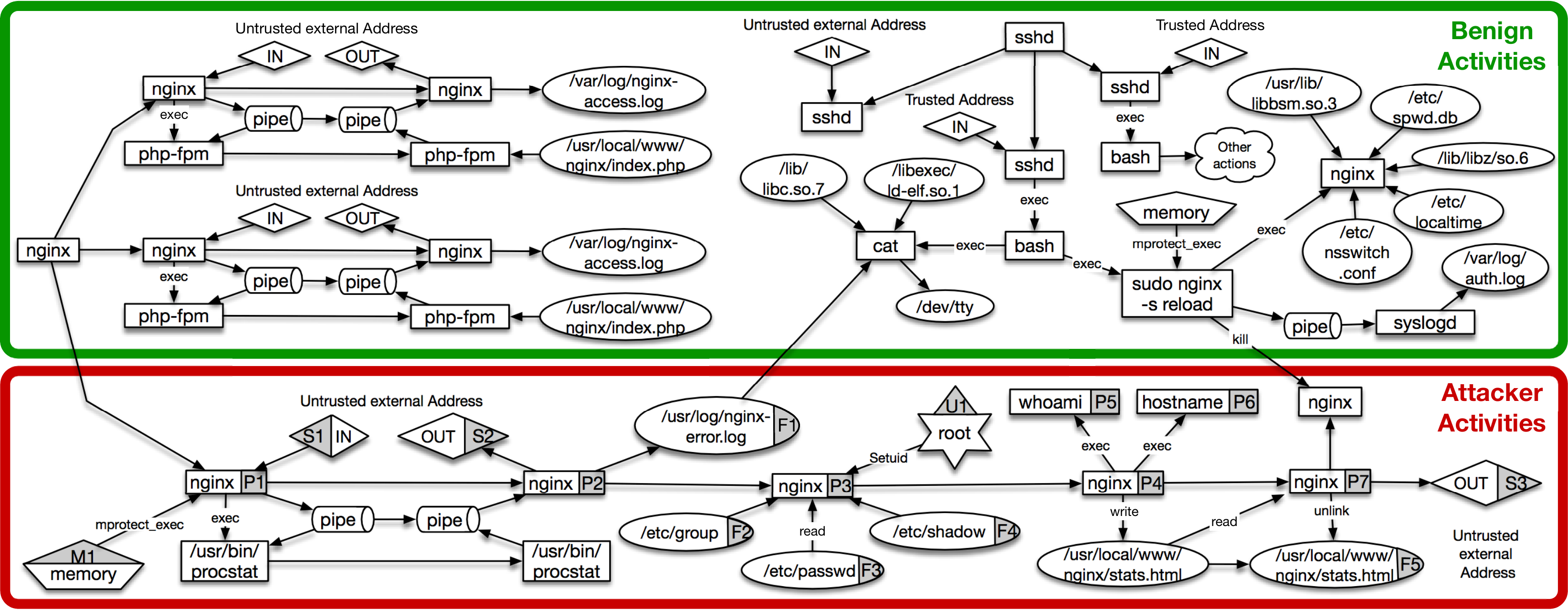}
  \end{center}
  \vspace*{-1em}
    \caption{Provenance Graph of the Running Example.}
    \vspace*{-1em}
    \label{fig:provgraph}
\end{figure*}

\noindent{\bf Evaluation.} We evaluated \projname on data generated by DARPA
Transparent Computing program that involved a professional red-team simulating
multiple cyber-attacks on a network consisting of different platforms. We
implemented appropriate system audit data parsers for Linux, FreeBSD, and
Windows, to process and convert their audit data to a common data representation
and analysis format. The advantage of using system audit data is that it is a
reliable source of information and is free of unauthorized tamper (under a
threat model of non-compromised kernel). 

Evaluation of \projname on nine
real-life APT attack scenarios, as well as running it as a real-time intrusion
detection tool in a live experiment spanning for two weeks, show that \projname
is able to clearly distinguish between attack and benign scenarios and can
discover cyber-attacks with high precision and recall (Sec. \ref{sec:eval}).

%% file: background.tex
\section{A Running Example} \label{running}
In this section, we present a running example used through the paper to
illustrate our approach. This example represents an attack carried out by a
red-team as part of a research program organized by a government agency
(specifically, US DARPA). In this attack, a vulnerable {\em Nginx}
web server runs on a {\em FreeBSD} system. Its operations (system calls)
are captured in the system audit log. From this audit data, we construct a {\em
  provenance graph}, a fragment of which is shown in Fig. \ref{fig:provgraph}.
Nodes in this graph represent system entities such as processes (represented as
rectangles), files (ovals), network connections (diamonds), memory objects
(pentagons), and users (stars). Edges correspond to system calls and are
oriented in the direction of information flow and/or causality. Note that our
provenance graph has been rendered acyclic using the (optimized) node versioning
technique described in Reference~\cite{depPresRed18}.

The goal of the attacker is to exfiltrate sensitive information from the system.
The attacker's activities are depicted at the bottom of Fig.
\ref{fig:provgraph}, and consist of the following steps:
\begin{itemize}
\item {\em Initial Compromise.} The attacker sends a malicious payload on the
  socket (S1) listening on port 80. As a result, {\em Nginx} makes some part of
  its memory region (M1) executable. Next, the attacker gains control over the
  {\em Nginx} process by using a reflective self-loading exploit.
\item {\em C\&C Communications.} The compromised {\em Nginx} process makes a connection (S2) to the C\&C server to receive commands from the attacker.
\item {\em Privilege Escalation.} The attacker exploits an existing vulnerability to escalate the privilege of {\em Nginx} to root (U1).
\item {\em Internal Reconnaissance.} Next, the attacker issues commands such as
  {\em whoami} (P5) and {\em hostname} (P6). These commands were used by the
  red team to simulate access to confidential/proprietary data. The attacker
  also reads usernames and password hashes (F2, F3, F4) and writes all this
  information to a temporary file.
\item{\em Exfiltration.} Next, the attacker transfers the file containing the gathered information to her/his machine (S3).
\item{\em Cleanup.} In the last step of the attack, the attacker removes the temporary file (F5) to clean up any attack remnants. 
\end{itemize}

\noindent This example illustrates many key challenges 
described below:

\noindent
{\bf Stealthy Attacks}. This attack leaves a minimal footprint on the system.
The first step of the attack, the initial compromise of the {\em Nginx} server,
is executed in main memory and does not leave any visible traces such as
downloaded files. Moreover, the payload runs within the existing \texttt{Nginx}
process. It is very challenging to detect such stealthy attacks, where attacker
activities blend in seamlessly with normal system operation.

\noindent
{\bf Needle in a haystack}. Even a single host can generate tens of millions of
events per day. All but a very tiny fraction of these --- typically much less
than 0.01\% --- correspond to benign activities. (The top portion of Fig.
\ref{fig:provgraph} shows a small subset of benign activities in the audit log.)
It is difficult to detect such rare events without a high rate of false alarms.
More importantly, it is very challenging to filter out these benign events from
the attack summaries presented to analysts.

\noindent
{\bf Real-time detection}. We envision \projname to be used in conjunction with
a {\em cyber-response} system, so it is necessary to detect and summarize an
ongoing campaign in a matter of seconds. Real-time detection poses additional
challenges and constraints for the techniques used in \projname.

To overcome these challenges, note that, despite blending seamlessly into
benign background activity, two factors stand out regarding the attack. First,
the attack steps achieve capabilities corresponding to some of the APT stages.
Second, the attack activities are connected via information flows.
In the next section, we describe the \projname approach based on
these two key observations.

%% file: approach.tex
\section{Approach Overview}\label{sec:approach}
The central insight behind our approach is that even though the concrete attack
steps may vary widely among different APTs, the high-level APT behavior often
conforms to the same kill-chain introduced in Section~\ref{sec:intro} (Figure
\ref{fig:apt-lifecycle}). Our analysis of hundreds of APT reports from
~\cite{apt-reports} suggests that most APTs consist of a subset, if not all, of
those steps. More importantly, we make the observation that these steps need to
be {\em causally connected}, and this connectedness is a major indication that
an attack is unfolding. 

Note that the concrete manifestation of each APT step may vary, e.g., an initial
compromise may be executed as a drive-by-download or as a spear-phishing attack
with a malicious file that is executed by a user. Regardless, the APT steps
themselves represent a high-level abstraction of the attacker's intentions, and
hence they must manifest themselves even if the operational tactics used by
attackers vary across APTs. Moreover, information flow or causal relations must
necessarily exist between them since the APT steps are logically dependent on
each other, e.g., exfiltration is dependent on internal reconnaissance to gather
sensitive data.

The research question, therefore, is whether we can base our detection on 
\begin{itemize}
\item an APT's most essential high-level behavioral steps, and 
\item the information flow dependencies between these steps. 
\end{itemize}
A major challenge in answering this question is the large semantic gap between
low-level audit data and the very high-level kill-chain view of attacker's
goals, intentions, and capabilities.

\begin{figure*}[ht!]
  \begin{center}
    \includegraphics[width=0.8\textwidth,height=1.74in]{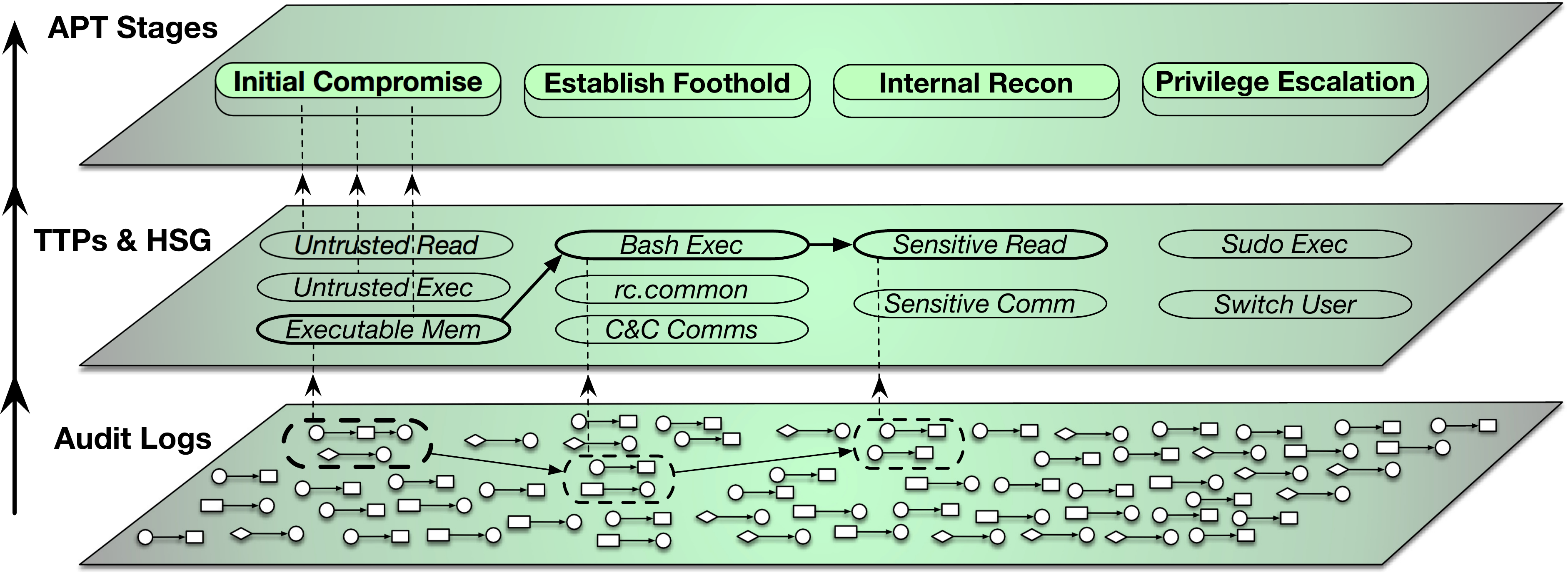}
  \end{center}
  \vspace*{-1em}
    \caption{\projname Approach: From Audit Records to High-Level APT Stages}
    \vspace*{-0.6em}
    \label{fig:approach}
\end{figure*}

\noindent
\textbf{Bridging the Semantic Gap}. To bridge the semantic gap between low-level
system-call view and the high-level kill-chain view, we build an intermediate
layer as shown in Fig. \ref{fig:approach}. The mapping to this intermediate
layer is based on MITRE's ATT\&CK framework \cite{attAndck}, which describes
close to 200 behavioral patterns defined as \textit{Tactics, Techniques, and
Procedures (TTPs)} observed in the wild.

Each TTP defines one possible way to realize a particular high-level capability.
For instance, the capability of \textit{persistence} in a compromised Linux
system can be achieved using 11 distinct TTPs, each of which represents a
possible sequence of lower level actions in the ATT\&CK framework, e.g.,
installation of a rootkit, modification of boot scripts, and so on. These lower
level actions are closer to the level of abstraction of audit logs, so it is
possible to describe TTPs in terms of nodes and edges in the provenance graph.

\noindent
\textbf{Technical challenges}.   The main technical challenges in
realizing the approach summarized in Fig. \ref{fig:approach} are:
\begin{itemize}
\item {\em efficient matching} of low-level event streams to TTPs,
\item {\em detecting correlation} between  attack steps, and
\item {\em reducing} false positives.
\end{itemize}
We solve these challenges through several design innovations. For efficient
matching, we use a representation of the audit logs as a directed provenance
graph (Section~\ref{sec:design}) in main memory, which allows for efficient
matching. This graph also encodes the information flow dependencies that exist
between system entities (such as processes and files). TTPs are specified as
patterns that leverage these dependencies. For instance, in order to match a
\textit{maintain persistence} TTP, an information flow dependency must exist
from a process matching an \textit{initial compromise} TTP to the
\textit{maintain persistence} TTP.

For detecting correlations between attack steps, we build a \textit{High-level
  Scenario Graph (HSG)} as an abstraction over the provenance graph. Each node
in the HSG represents a matched TTP, while the edges represent information flow
and causality dependencies among those matched TTPs. An HSG is illustrated in
the middle layer of Fig. \ref{fig:approach} by nodes and edges in boldface. (We
refer the reader to Fig.~\ref{fig:highlevelgraph} for the HSG of the running
example.) To determine the edges among nodes in the HSG, use the
\textit{prerequisite-consequence} patterns of among the TTPs and the APT stages.

To reduce the number of false positives (i.e., HSGs that do not represent
attacks), we use a combination of: (a) learning benign patterns that may produce
false positive TTPs and, (b) heuristics that assign weights to nodes and
paths in the graph based on their severity, so that the HSGs can be
ranked, and the highest-ranked HSGs presented to the analyst. 

In summary, the high-level phases of an APT are operationalized using a common
suite of tactics that can be observed from audit data. These observations
provide evidence that some malicious activity may be unfolding. The job of
\projname, then, is to collect pieces of evidence and infer the correlations
among them and use these correlations to map out the overall attack campaign.

%% file: design.tex
\section{System Design} \label{sec:design}

\input{policy_single_example.tex}
Like most previous works \cite{lee2013high, gehani2012spade,
  bates2015trustworthy, ma2016protracer} that rely on OS audit data, we consider
the OS kernel and the auditing engine as part of the trusted computing base
(TCB). In other words, attacks on the OS kernel, the auditing system and the
logs produced by it are outside the scope of our threat model. We also assume
that the system is benign at the outset, so the initial attack must originate
external to the enterprise, using means such as remote network access, removable
storage, etc.
\subsection{Data Collection and Representation}
Our system relies on audit logs retrieved from multiple hosts that may run
different operating systems (OSes).~\footnote{The design of \projname makes it possible to take additional inputs such as  events and alerts from a variety of IDS/IPS, but we do not discuss this aspect of the system further in paper.}  For Linux, the source of audit data is {\tt
  auditd}, while it is \texttt{dtrace} for BSD and ETW for Windows. This raw
audit data is collected and processed into an OS-neutral format. This is the
input format accepted by \projname. This input captures events relating to
principals (users), files (e.g., operations for I/O, file creation, ownership,
and permission), memory (e.g., mprotect and mmap) processes (e.g., creation, and
privilege change), and network connections. Although the default auditing
system incurs nontrivial overheads, recent research has shown that  overheads
can be made small \cite{bates2015trustworthy,pohly2012hi}.

The data is represented as a graph that we call the {\em provenance graph}.
The general structure of this graph is similar to that of many previous forensic
analysis works \cite{king2003backtracking, lee2013high, ma2016protracer}: the
nodes of the graph include subjects (processes) and objects (files, pipes,
sockets) and the edges denote the dependencies between these entities and are
annotated with event names. There are some important differences as well: our
subjects, as well as objects, are {\em versioned.} A new version of a node is
created before adding an incoming edge if this edge changes the existing
dependencies (i.e., the set of ancestor nodes) of the node. Versioning
enables optimizations that can prune away a large fraction of events in
the audit log without changing the results of forensic analysis 
 \cite{depPresRed18}. Moreover, this versioned graph is acyclic, which
can simplify many graph algorithms.

Another significant point about our provenance graph is that it is designed to
be stored in main memory. We have developed a highly compact provenance graph
representation in our previous work \cite{hossain2017sleuth, depPresRed18} that,
on average, required less than 5 bytes of main memory per event in the audit
log. This representation enables real-time consumption of events and graph
construction over prolonged periods of time. It is on this provenance graph that
our analysis queries for behavior that matches our TTP specifications.
\subsection{TTP Specification}
TTP specifications provide the mapping between low-level audit events and
high-level APT steps. Therefore, they are a central component of our approach.
In this subsection, we describe three key choices in the TTP design that enable
efficient and precise attack detection.

Recall that in our design, TTPs represent a layer of intermediate abstraction
between concrete audit logs and high-level APT steps. Specifically, we rely on
two main techniques to lift audit log data to this intermediate layer: (a) an
OS-neutral representation of security-relevant events in the form of the {\em
  provenance graph} and (b) use of {\em information flow} dependencies between
entities involved in the TTPs. Taken together, these techniques enable
high-level specifications of malicious behavior that are largely independent of
many TTP details such as the specific system calls used, names of malware,
intermediate files that were created and the programs used to create them, etc.
In this regard, our information flow based TTP specification approach is more
general than the use of {\em misuse specifications}
~\cite{kumar1995classification, porras1992penetration} from the IDS
literature. Use of information flow dependencies is crucial
in the detection of stealthy APTs that hide their activities by using benign
system processes to carry out their goals.

In addition to specifying the steps of a TTP, we need to capture its
prerequisites. Prerequisites not only help reduce false positives but also help
in understanding the role of a TTP in the larger context of an APT campaign. In
our TTP specifications, prerequisites take the form of causal relationships and
information flows between APT stages.

Finally, TTP matching needs to be efficient, and must not require expensive
techniques such as backtracking. We find that most TTPs can be modeled in our
framework using a single event, with additional preconditions on the subjects
and objects involved. 

An example of a TTP rule specification is shown in Table
\ref{tab:policyExample}, with additional rules appearing in Section
\ref{sec:implementation}. In Table \ref{tab:policyExample}, the first column
represents the APT stage, and the second column represents the associated TTP
name and the entities involved in the TTP. The third column specifies the event
family associated with the TTP. For ease of illustration, some of the specific
events included in this family are shown in the fourth column, but note that
they are not part of a TTP rule. (Event classes are defined once, and reused
across all TTP rules.)

The fifth column represents a severity level associated with each TTP. We use
this severity level to rank alarms raised by our system, prioritizing the most
severe alarms. Our current assignment of the severity levels is based on the
Common Attack Pattern Enumeration and Classification (CAPEC) list defined by
US-CERT and DHS with the collaboration of MITRE \cite{CAPEC} but can be
tailored to suit the needs of a particular enterprise. We also provide another
customization mechanism, whereby each severity level can be mapped to an
analyst-specified weight that reflects the relative importance of different APT
stages in a deployment context.

The last column specifies the prerequisites for the TTP rule to match. The
prerequisites can specify conditions on the parameters of the TTP being matched,
e.g., the socket parameter $S$ for the $Untrusted\_Read$ TTP on the first row.
Prerequisites can also contain conditions on previously matched TTPs and their
parameters. For instance, the prerequisite column of the $Make\_Mem\_Exec(P,M)$
TTP contains a condition $\exists \ Untrusted\_Read(?, P')$. This prerequisite
is satisfied only if an $Untrusted\_Read$ TTP has been matched for a process
$P'$ earlier, and if the processes involved in the two TTPs have a
$path\_factor$ (defined below) less than a specified threshold. 

Prerequisites can capture relations between the entities involved in two TTPs,
such as the parent-child relation on processes, or information flow between
files. They can also capture the condition that two TTPs share a common
parent. Using prerequisites, we are able to prune many false positives, i.e.,
benign activity resembling a TTP.

\subsection{HSG Construction}
\begin{figure}
  \begin{center}
    \includegraphics[width=\columnwidth]{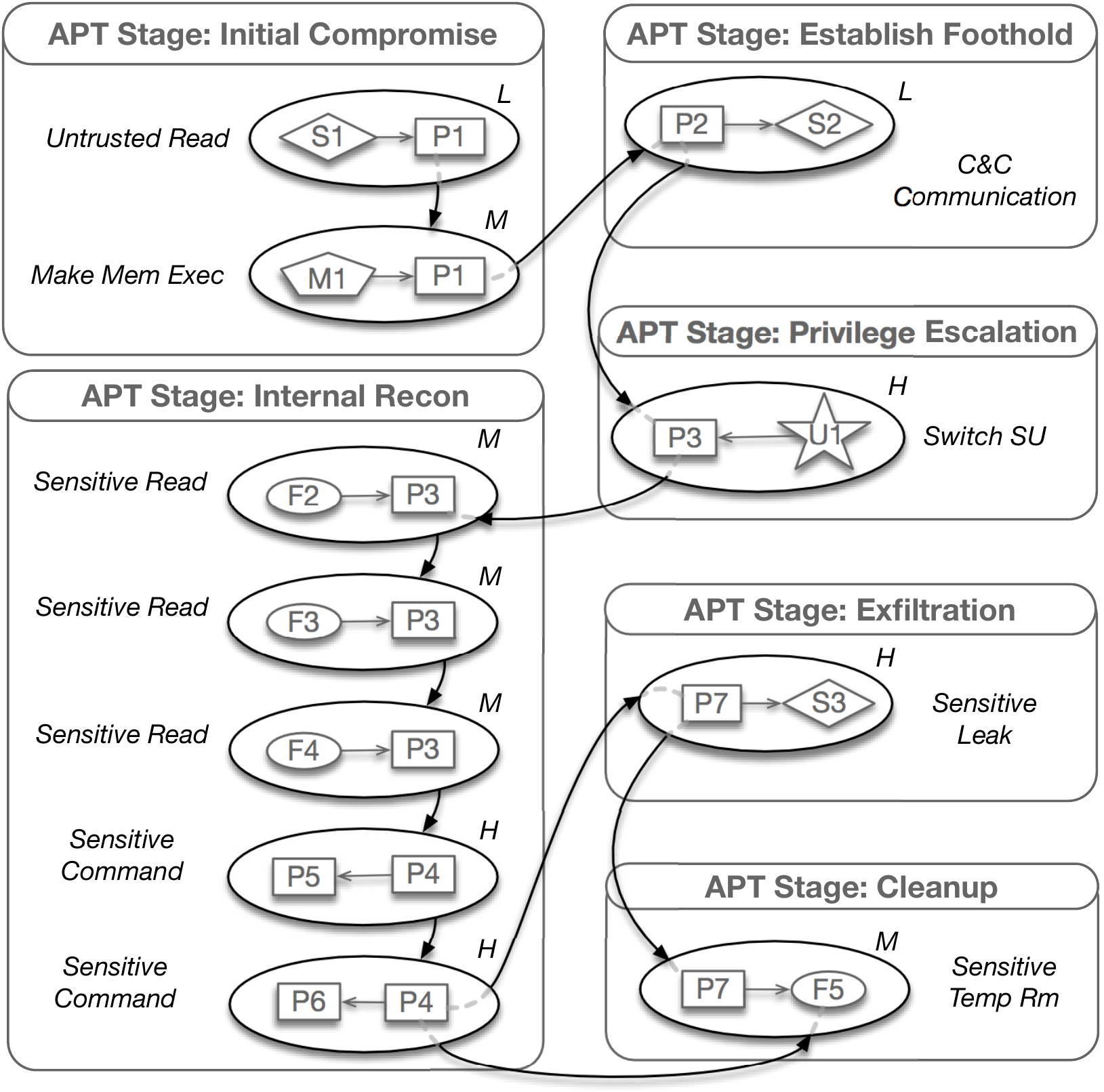}
  \end{center}
  \vspace*{-1em}
    \caption{High-Level Scenario Graph for the Running Example.}
    \vspace*{-1em}
    \label{fig:highlevelgraph}
\end{figure}

Fig.~\ref{fig:highlevelgraph} illustrates an HSG for the running example. The
nodes of this graph represent matched TTPs and are depicted by ovals in the
figure. Inside each oval, we represent the matched provenance graph entities in
grey. For illustration purposes, we have also included the name of the TTP, the
APT stage to which each TTP belongs, and the severity level (Low, Medium or
High) of each TTP. The edges of the graph represent the prerequisites between
different TTPs. The dotted lines that complete a path between two entities
represent the prerequisite conditions. For instance, the $Make\_Mem\_Exec$ TTP
has, as a prerequisite, an $Untrusted\_Read$ TTP, represented by the edge
between the two nodes.

The construction of the HSG is primarily driven by the {\em prerequisites:}
A TTP is matched and added to the HSG if all its prerequisites are satisfied.
This factor reduces the number of TTPs in the HSG at any time, making it possible
to carry out sophisticated analyses without impacting real-time performance.
\subsection{Avoiding Spurious Dependencies} \label{subsec:dep-expl}
By \textit{spurious dependencies}, we refer to uninteresting and/or irrelevant
dependencies on the attacker's activities. For instance, in
Fig.~\ref{fig:provgraph}, the process \mytt{nginx} (P2) writes to the file
\mytt{/usr/log/nginx-error.log}, and the \mytt{cat} process later reads that file.
However, even though there is a dependency between \mytt{cat} and the log file,
{\tt cat} is unrelated to the attack and is invoked independently
through {\tt ssh}. More generally, consider any process that consumes secondary
artifacts produced by the attack activity, e.g., a log rotation system that
copies a log file containing some fraction of entries produced by an attacker's
process. Such processes, although they represent benign background activity,
will be flagged in the provenance graph as having a dependence on the attacker's
processes. If these spurious dependencies aren't promptly pruned, there can be a
dependence explosion that can enormously increase the size of HSGs. As a result,
the final result presented to the analyst may be full of benign activities,
which can cause the analyst to miss key attack steps embedded in a large graph.
For this reason, we prioritize \textit{stronger} dependencies over
\textit{weaker} ones, pruning away the latter as much as possible.

Intuitively, we can say that a process $P_d$ has a {\em strong} dependency on a
process $P_a$ if $P_d$ is a descendant process of $P_a$. Similarly, a file or a
socket has a {\em strong} dependency on a process $P_a$ if $P_a$ or its
descendant processes write to this file/socket. More generally, consider two
entities and a path between them in the \textit{provenance graph} that indicates
an information between them. Determining if this flow represents a {\em strong}
or {\em weak} information flow is equivalent to determining if the entities in
the flow share \textit{compromised} ancestors. If they share
compromised ancestors, they are part of the attacker's activities, and there is
a {\em strong} dependency among them, which must be prioritized. Otherwise,
we consider the dependency to be weak and deemphasize it in our analysis.

To generalize the above discussion to a case where there may be multiple
compromised processes, we introduce the following notion of 
an {\em ancestral cover} $AC(f)$ of all processes on an information flow path
$f$:
\[
\forall p \in f \; \exists a \in AC(f)\;\;\mbox{$a=p$ or $a$ is an ancestor of $p$}
\]
Note that non-process nodes in $f$ don't affect the above definition. A {\em
  minimum ancestral cover}, $AC_{min}(f)$ is an ancestral cover of minimum size.
Intuitively, $AC_{min}(f)$ represents the minimum number of ancestors that an
attacker must compromise (i.e., the number of exploits) to have full control of
the information flow path $f$. For instance, consider again the flow from the
\mytt{nginx} process, which is under the control of the attacker, to the
\mytt{cat} process. Since these two processes share no common ancestors, the
minimum ancestral cover for the path among them has a size that is equal to 2.
Therefore, to control the \mytt{cat} process, an attacker would have to develop
an additional exploit for {\tt cat}. This requires the attacker to first find a
vulnerability in {\tt cat}, then create a corresponding exploit, and finally,
write this exploit into the log file. By preferring an ancestral cover of size 1,
we capture the fact that such an attack involving {\tt cat} is a lot less likely
than one where the attack activities are executed by \mytt{nginx} and its
descendants.

We can now define the notion of $path\_factor (N_1,N_2)$ mentioned
earlier in the discussion of TTPs. Intuitively, it captures the extent of the
attacker's control over the flow from $N_1$ to $N_2$. Based on the above
discussion of using minimum ancestral covers as a measure of dependency
strength, we define $path\_factor$ as follows. Consider all of the information
flow paths $f_1,...,f_n$ from $N_1$ to $N_2$, and let $m_i$ be the
minimum ancestral cover size for $f_i$. Then, $path\_factor(N_1,N_2)$ 
is simply the minimum value among $m_1,\ldots,m_n$. 

Note that if process $N_2$ is a child of $N_1$, then there is a path with just a
single edge between $N_1$ to $N_2$. The size of minimum ancestral cover for this
path is 1 since $N_1$ is an ancestor of $N_2$. In contrast, the (sole) path
from {\tt nginx} to {\tt cat} has a minimum ancestral cover of size 2, so
$path\_factor({\tt nginx}, {\tt cat})=2$.

We describe an efficient computation of $path\_factor$ in
Section~\ref{sec:implementation}. In our experience, the use of $path\_factor$
greatly mitigated dependency explosions by prioritizing attacker-influenced
flows. 
\subsection{Noise Reduction} \label{subsec:noise}
\begin{figure*}
  \begin{center}
    \includegraphics[width=\textwidth,height=1.45in]{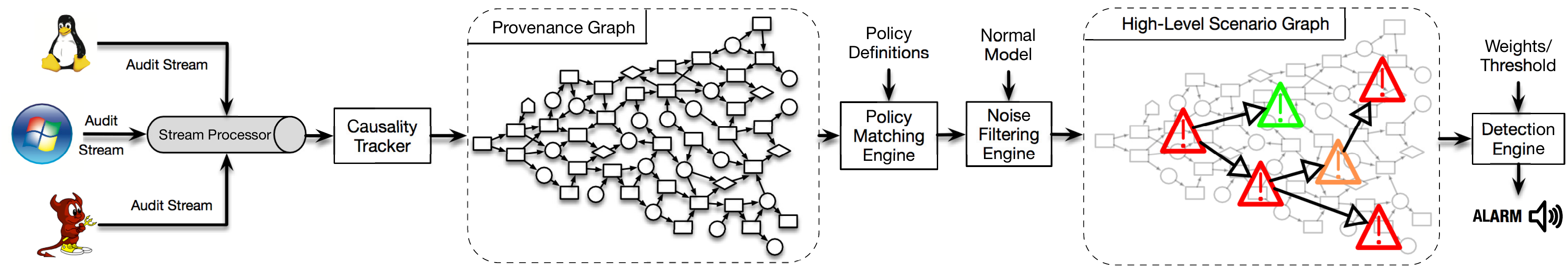}
  \end{center}
  \vspace*{-1em}
    \caption{\projname Architecture.}
    \vspace*{-1em}
    \label{fig:structure}
\end{figure*} 
One of the challenges in the analysis of audit logs for attack detection and
forensics is the presence of noise, i.e., benign events matching TTP rules.
Long-living processes such as browsers, web servers, and SSH daemons trigger TTP
matches from time to time. To cut down these false positives, we incorporate
noise reduction rules based on training data. We leverage two notions: (1)
benign prerequisite matches and (2) benign data flow quantity.

\noindent
{\bf Noise reduction based on benign prerequisites.} For each process, our
system learns prerequisites that fired frequently when the system is run
in a benign context. At runtime, when the prerequisites of a triggered TTP match
the prerequisites that were encountered during training, we ignore the match. 

\noindent
{\bf Noise reduction based on data flow quantity.} Filtering based on benign
prerequisites may lead to {\em false negatives:} a malicious event may go
unnoticed because it matches behavior observed during the learning phase. For
instance, even without any attack, \mytt{nginx} reads \texttt{/etc/passwd}
during its startup phase. However, if we were to whitelist all {\tt nginx}
access to {\tt /etc/passwd}, then a subsequent read by a compromised
\mytt{nginx} server will go unnoticed.

To tackle this problem, we enhance our learning to incorporate quantities of
information flow, measured in bytes transferred. For instance, the amount of
information that can flow from the file \mytt{/etc/passwd} to \mytt{nginx} is
equal to the size of that file, since \mytt{nginx} reads that file only once.
Therefore, if significantly more bytes are observed flowing from
\mytt{/etc/passwd} to \mytt{nginx}, then this flow \textit{may} be part of an
attack. To determine the cut-off points for information quantity, we observe
process-file and process-socket pairs over a period in a benign setting.
\subsection{Signal Correlation and Detection}\label{subsec:detection}
Given a set of HSGs, how do we distinguish the ones that constitute an attack
with a high confidence? We address this challenge by 
assigning a severity score to each HSG. This assignment proceeds in
two steps further described below. 

\noindent
\textbf{Threat Tuples}. First, we represent the attacker's progress in a
campaign by an abstract {\em threat tuple} associated with the corresponding
HSG. In particular, for every HSG, a {\em threat tuple} is a 7-tuple $\langle
S_{1}, S_{2}, S_{3},..., S_{7} \rangle$ where each $S_{i}$ corresponds to the
severity level of the APT stage at index $i$ of the HSG. We chose 7-tuples based
on an extensive survey of APTs in the wild~\cite{apt-reports}, but other choices
are possible as well. 

Since different TTPs belonging to a certain APT stage may have different
severity levels, there are usually multiple candidates to pick from. It is
natural to choose the highest severity level among these candidates. For
instance, the {\em threat tuple} associated with the HSG of Fig.
\ref{fig:highlevelgraph} is $\langle M,L,H,H,-,H,M \rangle$. This tuple contains
6 entries because its matched TTPs belong to 6 different APT stages. The entries
are ordered according to the order of the APT stages in the kill-chain. For
instance, the first entry of the tuple is M since the most severe TTP
belonging to \textit{Initial\_Reconnaissance} in the graph has severity~M.

\noindent 
\textbf{HSG Ranking and Prioritization}. To rank HSGs, we first transform a
threat tuple to a numeric value. In particular, we first map each element of a
threat tuple to a numerical value based on the conversion table
(Table~\ref{tab:NIST}) included in the Common Vulnerability Scoring System
(CVSS), a vendor-neutral industry standard created through the collaboration of
security professionals across commercial, non-commercial, and academic sectors~\cite{cvssScores}.
Alternative scoring choices may be made by an enterprise, taking into
context its perceived threats and past threat history.

\begin{table}[h]
 \begin{center}
   \begin{tabular}{|c|c|M{1.75cm}|}
      \hline
      \textbf{Qualitative level} & \textbf{Quantitative Range} & \textbf{Rounded up Average Value} \\
      \hline
      Low & 0.1 - 3.9 & 2.0 \\
      \hline
      Medium & 4.0 - 6.9 & 6.0 \\
      \hline
      High & 7.0 - 8.9 & 8.0 \\
      \hline
      Critical & 9.0 - 10.0 & 10.0 \\
      \hline                        
   \end{tabular}
 \end{center}

 \caption{NIST severity rating scale}\label{tab:NIST}
 \vspace*{-2em}
\end{table}

Next, we combine the numeric scores for the 7 APT stages into a single overall
score. The formula that we use to compute this score was designed with two main
criteria in mind: (1) flexibility and customization, and (2) the correlation of
APT steps is reflected in the magnification of the score as the steps unfold. To
address these criteria, we associate a weight with each entry in the converted
threat tuple and calculate a \textit{weighted product} of the threat tuple as
the score. These weights are configurable by a system administrator, and they
can be used to prioritize detection of specific stages over other stages. 

Using a training set, we performed several experiments and compared results
using other schemes, such as weighted sum, exponential sum, and geometric sum.
For each equation, we measured the average margin between the benign subgraph
scores and the attack subgraph scores after normalization and found that the
weighted product had the best results. Hence we use the following criteria to
flag an APT attack:
\begin{equation} \label{eq:1}
\prod_{i=1}^{n} (S_{i})^{w_{i}} \ge \tau
\end{equation}
Here, $n$ is the number of APT stages, $w_{i}$ and $S_{i}$ denote respectively
the weight and severity of stage $i$, and $\tau$ is the detection
threshold. If no TTP occurs in stage i, we set $S_{i} = 1$.

\input{policy_table}

%% file: policy_single_example.tex
\begin{table*}[t]
 \begin{center}
   \begin{tabular}{|m{2.2cm}|m{1.8cm}|c|m{3.2cm}|c|>{\raggedright\arraybackslash}m{5cm}|}
      \hline
      \textbf{APT Stage} & \textbf{TTP} & \textbf{Event Family} & \textbf{Events} & \textbf{Severity} & \textbf{Prerequisites} \\%& comments\\
      \hline
      $Initial\_$ $Compromise(P)$ & $Untrusted\_$ $Read (S,P)$ & READ & FileIoRead (Windows), read/pread/readv/preadv (Linux,BSD) & L & $S.ip \notin$ \{Trusted\_IP\_Addresses\} \\%& Entry point of attacker \\
      \cline{2-6}
       & $Make\_Mem\_$ $Exec(P,M)$ & MPROTECT & VirtualAlloc (Windows), mprotect (Linux,BSD) & M & \$PROT\_EXEC\$ $ \in M.flags$  $\land \ \exists \ Untrusted\_Read (?,P^\prime):path\_factor(P^\prime, P) <= path\_thres$ \\%& in-memory executions\\
      \hline
      $Establish\_$ $Foothold(P)$& $Shell\_$ $Exec(F,P)$ & EXEC  & ProcessStart (Windows), execve/fexecve (Linux,BSD)& M & $F.path \in$ \{Command\_Line\_Utilities\} $\land \ \exists \ Initial\_Compromise (P^\prime) : path\_factor(P^\prime, P) <= path\_thres$ \\%& attackers usually execute shell after the successful exploit\\  
      \hline
   \end{tabular}
 \end{center}
 \caption{Example TTPs. In the Severity column, L=Low, M=Moderate, H=High, C=Critical. Entity types are shown by the characters: P=Process, F=File, S=Socket, M=Memory, U=User.}
 \vspace*{-3em}
 \label{tab:policyExample}
\end{table*}

%% file: policy_table.tex
\begin{table*}[t]
 \begin{center}

   \begin{tabular}{|p{2cm}|c|>{\centering\arraybackslash}m{1.2cm}|c|>{\raggedright\arraybackslash}p{8.5cm}|}
      \hline
      \textbf{APT Stage} & \textbf{TTP} & \textbf{Event Family} & \textbf{Severity} & \textbf{Prerequisites} \\%& comments\\
      \hline
      $Initial\_$ $Compromise(P)$ & $Untrusted\_Read (S,P)$ & READ & L & $S.ip \notin$ \{Trusted\_IP\_Addresses\} \\%& Entry point of attacker \\
      \cline{2-5}
       & $Make\_Mem\_Exec(P,M)$ & MPROTECT & M & \$PROT\_EXEC\$ $ \in M.flags$  $\land \ \exists \ Untrusted\_Read (?,P^\prime):path\_factor(P^\prime, P) <= path\_thres$ \\%& in-memory executions\\
      \cline{2-5}
       & $Make\_File\_Exec(P,F)$ & CHMOD & H & \$PROT\_EXEC\$ $ \in F.mode$  $\land \ \exists \ Untrusted\_Read (?,P^\prime):path\_factor(P^\prime, F) <= path\_thres$ $\land \ \exists \ Untrusted\_Read (?,P^{\prime\prime}):path\_factor(P^{\prime\prime}, P) <= path\_thres$ \\%& in-memory executions\\
      \cline{2-5}      
       & $Untrusted\_File\_Exec(F,P)$ & EXEC & C & $\exists \ Untrusted\_Read (?,P^\prime) : path\_factor(P^\prime, F) <= path\_thres$ \\%& Executing an untrusted file\\
      \hline
      $Establish\_$ $Foothold(P)$ & $Shell\_Exec (F,P)$ & EXEC & M & $F.path \in$ \{Command\_Line\_Utilities\} $ \land \ \exists \ Initial\_Compromise (P^\prime) : path\_factor(P^\prime, P) <= path\_thres$ \\%& attackers usually execute shell after the successful exploit\\ 
      \cline{2-5}      
       & $CnC(P,S)$ & SEND & L &  $S.ip \notin$ \{Trusted\_IP\_Addresses\}$ \ \land \ \exists \ Initial\_Compromise (P^\prime) : path\_factor(P^\prime, P) <= path\_thres$\\%& connecting to mothership for receiving commands\\
      \hline
      $Privilege\_$ $Escalation(P)$ & $Sudo\_Exec (F,P)$ & EXEC & H & $F.path \in $ \{SuperUser\_Tools\}$ \ \land \ \exists \ Initial\_Compromise (P^\prime) : path\_factor(P^\prime, P) <= path\_thres$\\%& \\
      \cline{2-5}      
       & $Switch\_SU(U,P)$ & SETUID & H & $U.id \in $ \{SuperUser\_Group\}$ \ \land \ \exists \ Initial\_Compromise (P^\prime) : path\_factor(P^\prime, P) <= path\_thres$\\%& Gaining root priviledges\\
      \hline  
      $Internal\_$ $Recon(P)$ & $Sensitive\_Read (F,P)$ & READ & M & $F.path \in$ \{Sensitive\_Files\} $ \land \ \exists \ Initial\_Compromise (P^\prime) : path\_factor(P^\prime, P) <= path\_thres$\\%& local discovery by reading sensitive files, like /etc/passwd, /etc/shadow\\
      \cline{2-5}      
       & $Sensitive\_Command(P,P^{\prime})$ & FORK & H & $P^{\prime}.name \in$ \{Sensitive\_Commands\} $\land \ \exists \ Initial\_Compromise (P^{\prime\prime}) : path\_factor(P^{\prime\prime}, P) <= path\_thres$\\%& commands like whoami, hostname, netstat\\
      \hline            
      $Move\_$ $Laterally(P)$& $Send\_Internal (P,S)$ & SEND & M & $S.ip \in$ \{Internal\_IP\_Range\} $\land \ \exists \ Initial\_Compromise (P^\prime) : path\_factor(P^\prime, P) <= path\_thres$\\%& Scanning Activity \\
      \hline   
      $Complete\_$ $Mission(P)$ & $Sensitive\_Leak (P,S)$ & SEND & H & $S.ip \notin$ \{Trusted\_IP\_Addresses\} $ \land \ \exists \ Internal\_Reconnaissance (P^\prime) : path\_factor(P^\prime, P) <= path\_thres$ $\land \ \exists \ Initial\_Compromise (P^{\prime\prime}) : path\_factor(P^{\prime\prime}, P) <= path\_thres$\\%& Exfiltrating sensitive 
      \cline{2-5}      
       & $Destroy\_System(F,P)$ & WRITE/ UNLINK & C & $F.path \in$ \{System\_Critical\_Files\} $\land \ \exists \ Initial\_Compromise (P^\prime) : path\_factor(P^\prime, P) <= path\_thres$\\%& disrupting normal operations of system\\            
      \hline  
      $Cleanup\_$ $Tracks(P)$ & $Clear\_Logs(P,F)$ & UNLINK & H & $F.path \in $ \{Log\_Files\}$ \land \ \exists \ Initial\_Compromise (P^\prime) : path\_factor(P^\prime, P) <= path\_thres$\\%& Attacker removes track of his activities\\        
      \cline{2-5}      
       & $Sensitive\_Temp\_RM(P,F)$ & UNLINK & M & $\exists \ Internal\_Reconnaissance (P^\prime) : path\_factor(P^\prime, F) <= path\_thres$ $\land \ \exists \ Initial\_Compromise (P^{\prime\prime}) : path\_factor(P^{\prime\prime}, P) <= path\_thres$\\%& Attackers compress/encrypt all the sensitive data to a file and remove that file after exfiltration\\ exfiltration\\             
      \cline{2-5}      
       & $Untrusted\_File\_RM(P,F)$ & UNLINK & M & $\exists \ Initial\_Compromise (P^\prime) : path\_factor(P^\prime, F) <= path\_thres$ $\land \ \exists \ Initial\_Compromise (P^{\prime\prime}) : path\_factor(P^{\prime\prime}, P) <= path\_thres$\\%& Attacker finally removes any executable file that is dropped as part of his malicious activities\\      
      \hline  

   \end{tabular}
 \end{center}
 \caption{Representative TTPs. Event family denotes a set of corresponding events in Windows, Linux, and FreeBSD. In the Severity column, L=Low, M=Moderate, H=High, C=Critical. Entity types are shown by the characters: P=Process, F=File, S=Socket, M=Memory, U=User.}
 \vspace*{-2em}
 \label{tab:policies}
\end{table*}

%% file: implementation.tex
\section{Implementation} \label{sec:implementation}
\noindent
{\bf Stream Consumption for Provenance Graph Construction}. 
Fig. \ref{fig:structure}
shows the architecture of \projname. To achieve platform independence, audit
records from different OSs are normalized to a common data representation (CDR)
with shared abstractions for various system entities. For streamlined audit data
processing, CDR-based audit records are published to a stream processing server
({\em Kafka}) and real-time analysis and detection proceeds by consuming from
the streaming server. We use our {\sc Sleuth} system \cite{hossain2017sleuth}
for stream consumption, causality tracking, and provenance graph construction, so
we don't describe those steps in detail here.

\noindent
\textbf{Policy Matching Engine and HSG Construction}. The \textit{Policy
  Matching Engine} takes the TTP rule specifications as input and operates on
the provenance graph. A representative set of the TTP rule specifications used
in the current implementation of \projname is shown in Table \ref{tab:policies}.
To match a TTP, as the provenance graph is being built, the policy matching
engine iterates over each rule in the rules table and its prerequisites. A
particularly challenging part of this task is to check, for each TTP,
the prerequisite conditions about previously matched TTPs and the
\textit{path\_factor}. In fact, previously matched TTPs may be located in a
distant region of the graph and the \textit{path\_factor} value may depend on
long paths, which must be traversed. We note that a common practice in prior
work \cite{king2005enriching,hossain2017sleuth,ma2016protracer,lee2013high} on
attack forensics is to do backward tracking from a TTP matching point to reach
an initial compromise point. Unfortunately, this is a computationally expensive
strategy in a real-time setting as the provenance graph might contain millions
of events.

To solve this challenge without backtracking, we use an incremental matching
approach that stores the results of the previous computations and matches and
propagates pointers to those results along the graph. When a specific TTP, which
may appear as a prerequisite condition in other TTPs, is matched, we create the
corresponding node in the HSG and a pointer to that node. The pointer is next
propagated to all the low-level entities that have dependencies on the entities
of that matched TTP.

The \textit{path\_factor} is similarly computed. In particular, given a matched
TTP represented as a node in the HSG, a \textit{path\_factor} value is
incrementally computed for the nodes of the provenance graph that have
dependencies on the entities of the matched TTP.
Assuming $N_1$ as a process generating an event matching a TTP,
$path\_factor(N_1,N_1)$ is initially assigned to 1.
Subsequently, when an edge $(N_1,N_2)$ is added to the provenance graph, $path\_factor(N_1,N_2)$ will be $1$ if $N_2$ is a non-process node or if it is a process with at least one common ancestor with $N_1$. Otherwise, the $path\_factor$ value increases by $1$.
In cases that an information flow happens from $N_2$ to $N_3$ while both $N_2$ and $N_3$ already have a dependency flow from $N_1$, a new version of $N_3$ is constructed, and the $path\_factor(N_1,N_{3\_new})$ is set to the minimum among the $path\_factors$ calculated by both flows. Note that in the acyclic provenance graph which is built based on this versioning system, the $path\_factor(N_1,N_2)$ never changes once it is set.
Finally, when an event
corresponding to a TTP event is encountered, we can reuse the pointer to the
prerequisite TTPs and the precomputed \textit{path\_factor} immediately if they
are available.

An expected bottleneck for this pointer-based correlation of the two layers
(provenance graph and HSG) is the space overhead and complexity it adds as the
provenance graph grows over time. Our operational observation is that,
typically, a large number of entities point to the same set of TTPs; This
phenomenon is not random and is actually the result of the propagation of pointers
in the process tree, from parent processes to all their descendants. It is, in
fact, rare that new pointers get added as the analysis proceeds. In general, the
key implementation insight is to maintain an intermediate object that maps
entities of the provenance graph to TTPs of the HSG. Therefore, each entity in
the provenance graph has only one pointer pointing to the intermediate mapper,
and the mapper object contains the set of pointers.

\begin{figure*}[t]
  \begin{center}
    \includegraphics[width=.7\columnwidth]{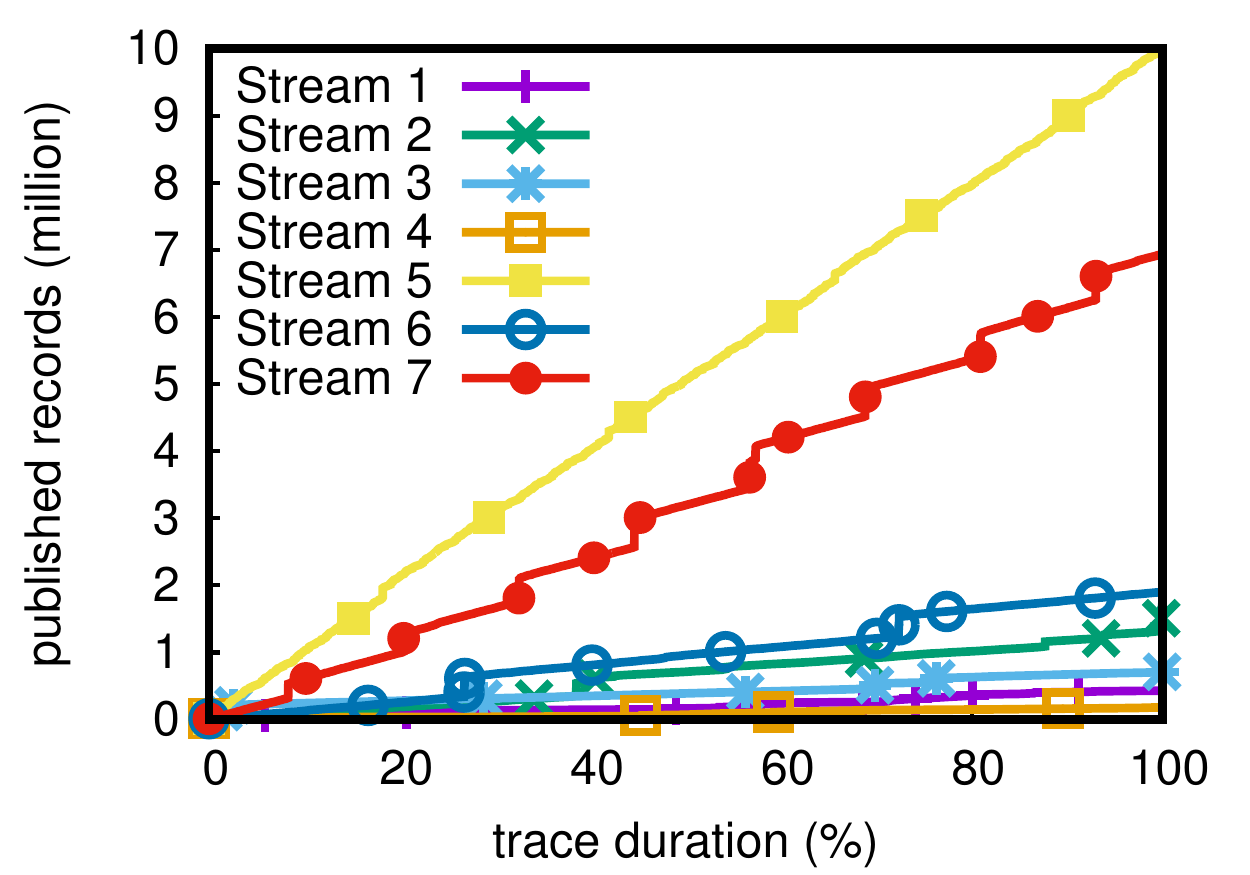}
    \hspace*{0.2\columnwidth}
    \includegraphics[width=.7\columnwidth]{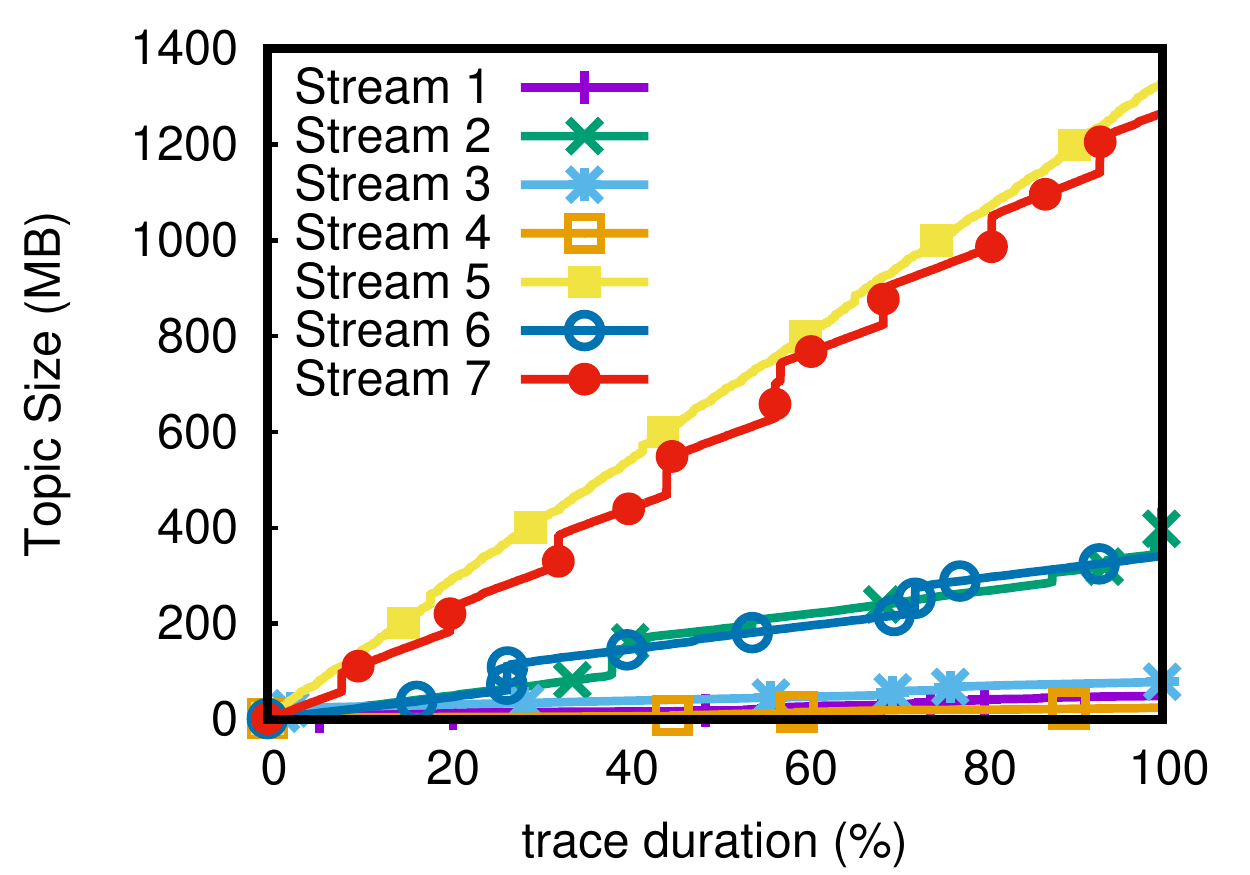}
  \end{center}
  \vspace*{-1.3em}
    \caption{(Left): Number (millions) of published records vs. \% of trace duration. (Right): Topic Size (MB) vs. \% of trace duration.}\label{fig:records_per_time}
   \vspace*{-1em}
\end{figure*} 

\noindent
\textbf{Noise Filtering and Detection Engines}. The \textit{Noise Filtering
  Engine} identifies benign TTP matches so that they can be excluded from the
HSG. It takes as input the normal behavior model learned on benign runs. This
model contains a map of the TTPs that are matched in benign runs and the
threshold on the number of bytes read from or written to system objects on these
runs. When the policy matching engine matches a new TTP, the entities and
prerequisites of that TTP are searched in this model. If an
entry exists in the model that contains all the prerequisites and the matched
event (having the same entity names), then the total amount of transferred bytes
is checked against the benign threshold. If the total amount of bytes
transferred is lower than the benign threshold, then the node corresponding to
the matched TTP is filtered out; otherwise, a node corresponding to it gets
added to the HSG. Finally, the {\em detection engine} computes the weighted sums
of the different HSGs and raises alarms when that value surpasses the detection
threshold.

%% file: eval.tex
\section{Experimental Evaluation}
\label{sec:eval}
\begin{table}[b]
 \begin{center}
   \begin{tabular}{|M{0.7cm}|c|M{1.4cm}|M{0.8cm}|M{1.1cm}|M{1.2cm}|}
      \hline
      \textbf{Stream No.} & \textbf{Duration} & \textbf{Platform} & \textbf{Scenario No.} & \textbf{Scenario Name} & \textbf{Attack Surface}\\
      \hline
      1 & 0d1h17m  & Ubuntu 14.04 (64bit) & 1 & Drive-by Download & Firefox 42.0\\
      \hline
      2 & 2d5h8m  & Ubuntu 12.04 (64bit) & 2 &  Trojan & \multirow{2}{*}{Firefox 20.0}\\
      \hline 
      3 & 1d7h25m  & Ubuntu 12.04 (64bit) & 3  &  Trojan & \multirow{2}{*}{Firefox 20.0} \\
      \hline \hline 
      4 & 0d1h39m & Windows 7 Pro (64bit) & 4 &  Spyware & Firefox 44.0  \\
      \hline       
      5 & 5d5h17m & Windows 7 Pro (64bit) & 5.1  &  Eternal Blue & Vulnerable SMB \\
      \cline{4-6}  
       & & & 5.2 &   RAT & Firefox 44.0 \\    
      \hline    \hline 
      6 & 2d5h17m & FreeBSD 11.0 (64bit) & 6  &  Web-Shell& Backdoored Nginx \\
      \hline       
      7 & 8d7h15m & FreeBSD 11.0 (64bit) & 7.1 &  RAT & Backdoored Nginx  \\
      \cline{4-6}  
       & & & 7.2 & Password Hijacking & Backdoored Nginx \\
      \hline  
   \end{tabular}
 \end{center}

 \caption{Datasets. Streams 5 and 7 contain two independent attack vectors occurring on the same host.}\label{datasets}
 \vspace*{-2em}
\end{table}

Our experimental evaluation is done on red-team vs. blue-team adversarial engagements organized by a government agency (specifically, US DARPA). We first evaluated \projname on a dataset that was available to us beforehand (Sections \ref{sec:eval:dataset},\ref{sec:eval:setup},\ref{sec:eval:results},\ref{sec:eval:scenarios}). Using this evaluation, we calculate the optimal threshold value for \projname in Section \ref{sec:eval-optimal-val}, and measure its performance in Section \ref{sec:eval:measure}. Finally, in Section \ref{sec:eval-live}, we explored applicability of \projname as a real-world live detection system in a setting that we have no prior knowledge of when or how red-team is conducting the attacks. After our live experiment, this dataset has been released in the public domain \cite{tc_github} to stimulate further research in this area.

\subsection{Datasets}\label{sec:eval:dataset}
\noindent
{\bf Attacks.}
The datasets we used for evaluation are summarized in Table \ref{datasets}. This table shows nine APT scenarios from 7 hosts across three OS platforms. There are three scenarios for each platform. Collectively, the streams cover 20 days' worth of audit logs collected using {\em auditd}, {\em dtrace}, and {\em ETW} from Ubuntu, FreeBSD, and Microsoft Windows, respectively. Each stream contains kernel audit logs of routine system activities and attack activities. Attacks constitute less than 0.001\% of the audit data volume. Streams 5 and 7 each contain two independent APT attacks, while the remaining streams contain one
APT attack each. 

In a nutshell, the adversarial goals and activities in the red team attack
scenarios cover those of high-profile APT campaigns. These include
typical APT activities such as browser-induced drive-by initial compromises,
backdoor injection, privilege escalation, internal reconnaissance, exfiltration
of sensitive assets, and cleanup of attack footprints. In these attacks,
sophisticated attack vectors such as reflective loading, web-shell capabilities,
and in-memory module loading were used by a red-team to evaluate the robustness
of \projname against APT attacks.

\noindent
{\bf Benign data.}
To build noise reduction models%
, we used $4$ days' worth of benign audit data streamed from the target hosts. %
These programs include browsers (e.g., Firefox), web servers (e.g., {\em Nginx}), and various daemons (e.g., {\em postfix}, {\em syslogd}).

\subsection{Setup}\label{sec:eval:setup}

Before the engagement, each machine is set up with some vulnerable software that later gets exploited. To further mix normal and attack logs, the red-team also performed benign activities on the target hosts in parallel with attacks. Routine system activities include, but are not limited to, web browsing, checking email, software upgrading, administrative tasks using PowerShell (in Windows), running programs that require administrative privileges, and so on.
\projname subscribed to the 7 Kafka topics (one topic per host) to consume audit streams and perform analysis and detection in real-time. The number of records published (in millions) with regards to the percentage growth of the audit trace duration is shown in Fig. \ref{fig:records_per_time} (left). Note that \projname consumes as fast as the publishing rate from the Kafka server. Fig. \ref{fig:records_per_time} (right) shows the incremental growth in the size of records published into each Kafka topic.

We configured \projname with TTPs mentioned in Table \ref{tab:policies} and set {\em path\_thres} $=3$ for prerequisites on TTPs and {\em weight} $= (10+i)/10$ for APT stage $i$, which takes into account slightly higher weights for later APT stages.

\subsection{Results in a Nutshell} \label{sec:eval:results}
Table \ref{results} summarizes the detection of the nine attack scenarios. The second column shows the {\em threat tuple} of each HSG matched during detection, and the third column shows the corresponding {\em threat score}. 
The fourth column shows the highest score among all benign scenarios of the machine on which the attack scenario is exercised. These benign scenarios might contain the exact programs in the corresponding attack scenario. 

The highest score assigned to benign HSGs is 338 (Scenario-3), and the lowest score assigned to attack HSGs is 608 (Scenario-5.2)
which is related to an incomplete attack with no harm done to the system. 
This shows that \projname has separated attack and benign scenarios into two disjoint clusters, and makes a clear distinction between them.

\begin{table}[t]
 \begin{center}
   \begin{tabular}{|M{.8cm}|c|M{.8cm}|M{2cm}|}
      \hline
      \textbf{Scenario No.} & \textbf{Threat Tuple} & \textbf{Threat Score} & \textbf{Highest Benign Score in Dataset} \\
      \hline
      1 &$\langle C,M,-,H,-,H,M\rangle$ & 1163881 & 61 \\
      \hline  
      2 & $\langle C,M,-,H,-,H,-\rangle$ & 55342 & 226 \\
      \hline
      3 &$\langle C,M,-,H,-,H,M\rangle$ & 1163881 & 338 \\
      \hline         
      4 & $\langle C,M,-,H,-,-,M\rangle$ & 41780 & 5 \\
      \hline
      5.1 & $\langle C,L,-,M,-,H,H\rangle$ & 339504 & 104\\
      \cline{1-3} 
      5.2 & $\langle C,L,-,-,-,-,M\rangle$ & 608 &  \\
      \hline         
      6 & $\langle L,L,H,M,-,H,-\rangle$ & 25162 & 137 \\
      \hline
      7.1 & $\langle C,L,H,H,-,H,M\rangle$ & 4649220 & 133\\
      \cline{1-3} 
      7.2 & $\langle M,L,H,H,-,H,M\rangle$ & 2650614 &  \\
      \hline
   \end{tabular}
 \end{center}
 \caption{Scores Assigned to Attack Scenarios.  L = Low, M = Moderate, H = High, C = Critical. {\bf Note:} for each scenario,  Highest Benign Score in Dataset is the highest {\em threat score} assigned to benign background activities streamed during the audit log collection of a host (pre-attack, in parallel to attack, and post-attack).}\label{results}
 \vspace*{-3em}
\end{table}

\begin{figure}[b]
  \begin{center}

    \includegraphics[width=.8\columnwidth]{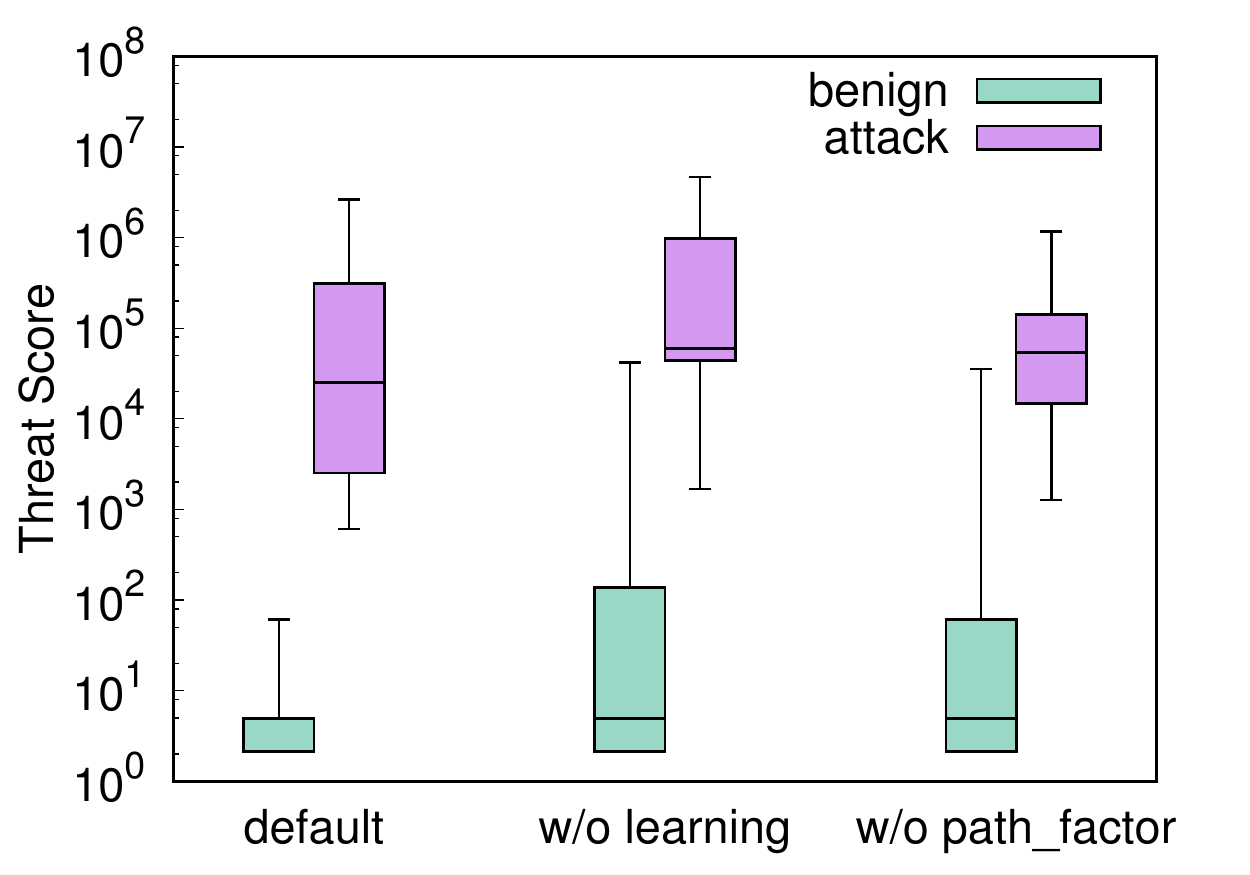}
  \end{center}
  \vspace*{-1em}
    \caption{Effects of Learning and {\em path\_factor} on Noise Reduction. Box covers from first to third quartiles while a bar in the middle indicates median, and whisker is extended from 10th to 90th percentile.}%
    \vspace*{-1em}
    \label{fig:learning}
\end{figure}

The effect of learning noise reduction rules and {\em path\_factor}  are shown in Fig. \ref{fig:learning}. This plot shows {\em threat score} for all benign and attack HSGs which are constructed after analyzing all the seven streams. These scores are shown under three different settings: default which both learning and {\em path\_factor} calculations are enabled, without learning, and without {\em path\_factor}.
It is obvious in the figure that with learning and {\em path\_factor}, there is  a more considerable margin between attack HSGs and benign ones.
Without learning or {\em path\_factor}, we notice an increase in noise, which leads to  false positives or false negatives. The 10th percentile, first quartile, and median of default box are all colliding on the bottom line of this box (score$=2.1$). This means that more than 50\% of {\em threat scores} are $2.1$, which is the result of having many HSGs with only one low severity {\em Untrusted Read} TTP.

\begin{figure}[t]
  \begin{center}

    \includegraphics[width=\columnwidth]{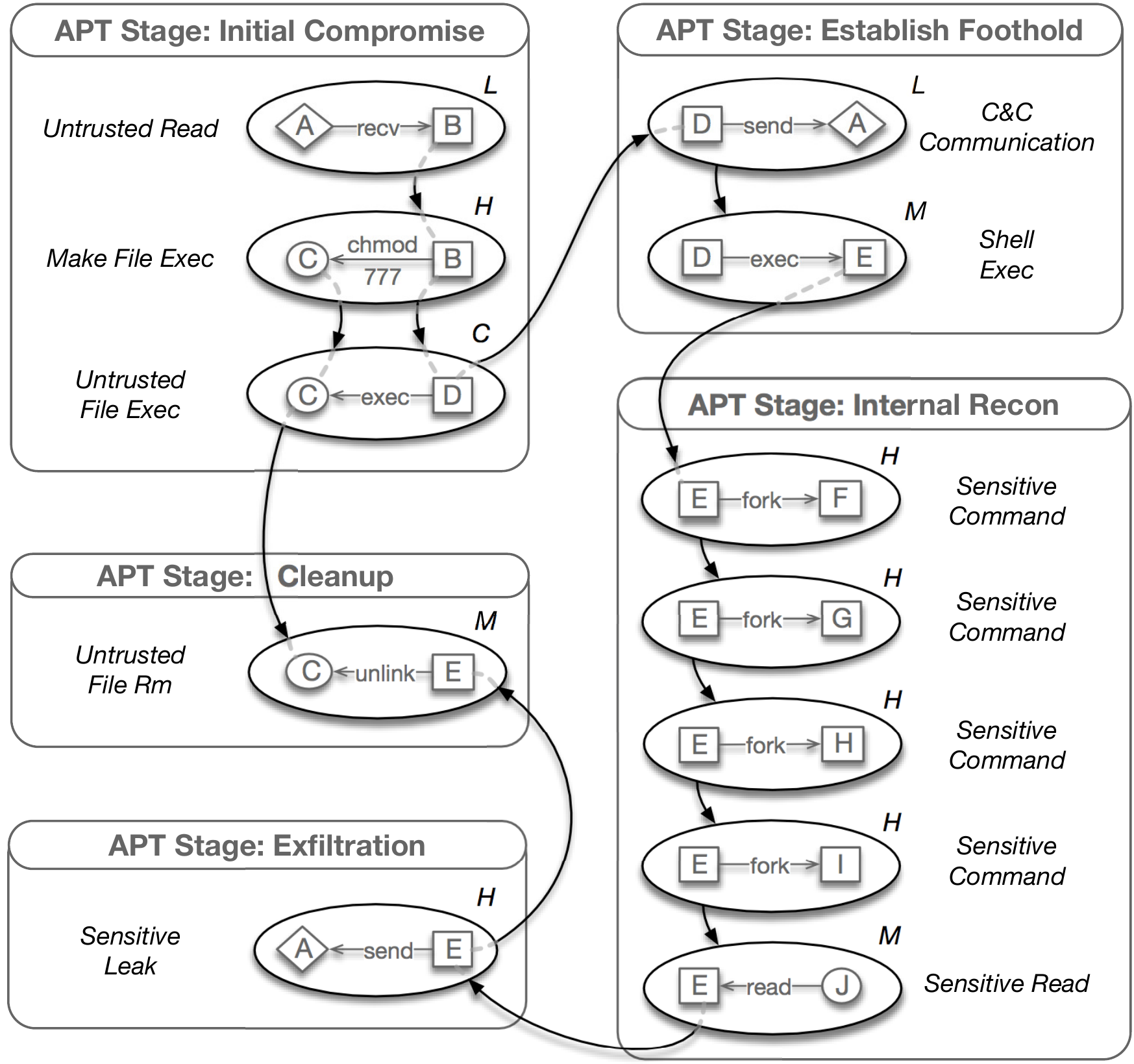}
  \end{center}
  \vspace*{-1em}
    \caption{HSG of Scenario-1 (Drive-by Download). Notations: A= Untrusted External Address; B= Firefox; C= Malicious dropped file (net); D= RAT process; E= bash; F= whoami; G= uname; I= netstat; J= company\_secret.txt;}
    \vspace*{-1em}
    \label{fig:drive-by-hsg}
\end{figure} 

\begin{figure}[b]
  \begin{center}

    \includegraphics[width=.8\columnwidth]{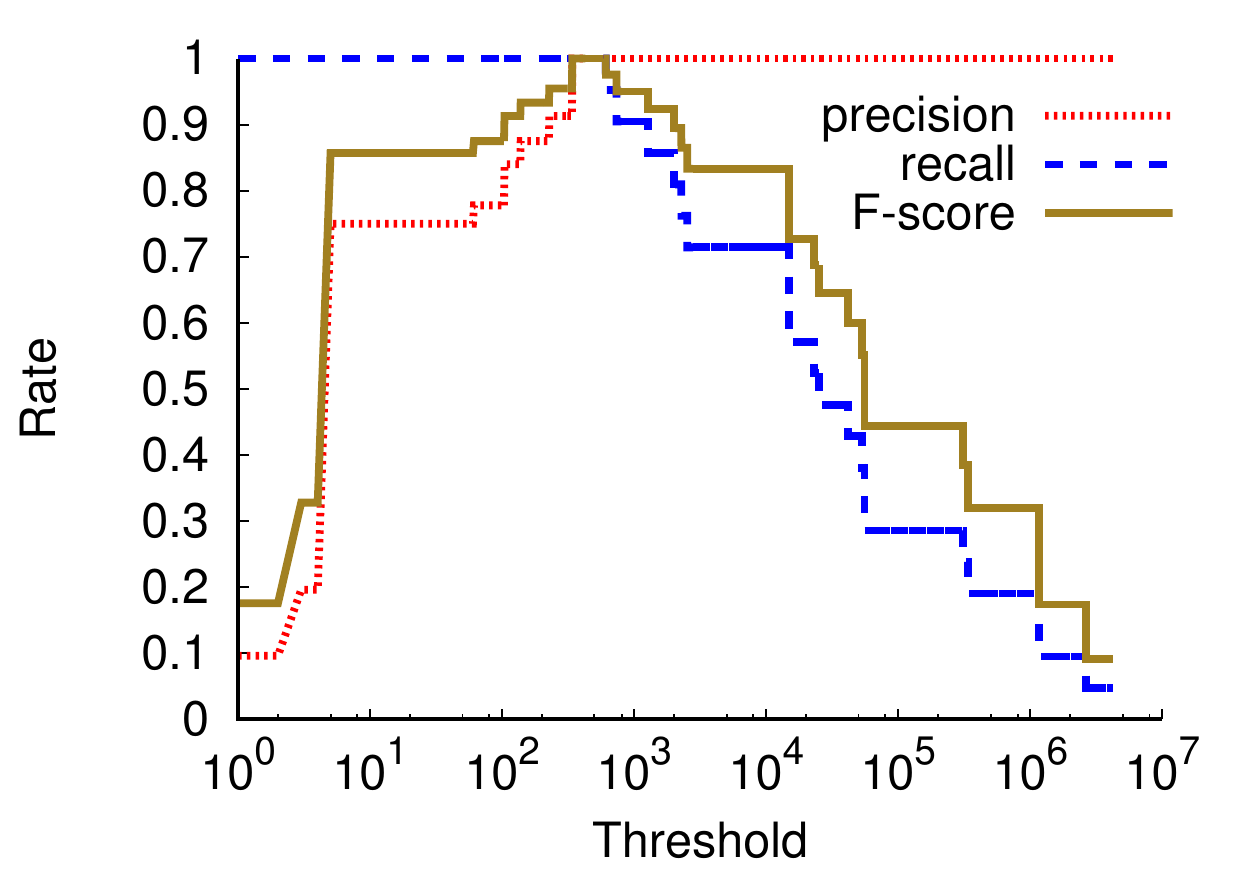}
  \end{center}
  \vspace*{-1em}
    \caption{Precision, Recall, and F-score of attack detection by varying the threshold value.}
    \vspace*{-.5em}
    \label{fig:fscore}
\end{figure} 

\begin{figure*}[!ht]
  \begin{center}

    \includegraphics[width=.7\columnwidth]{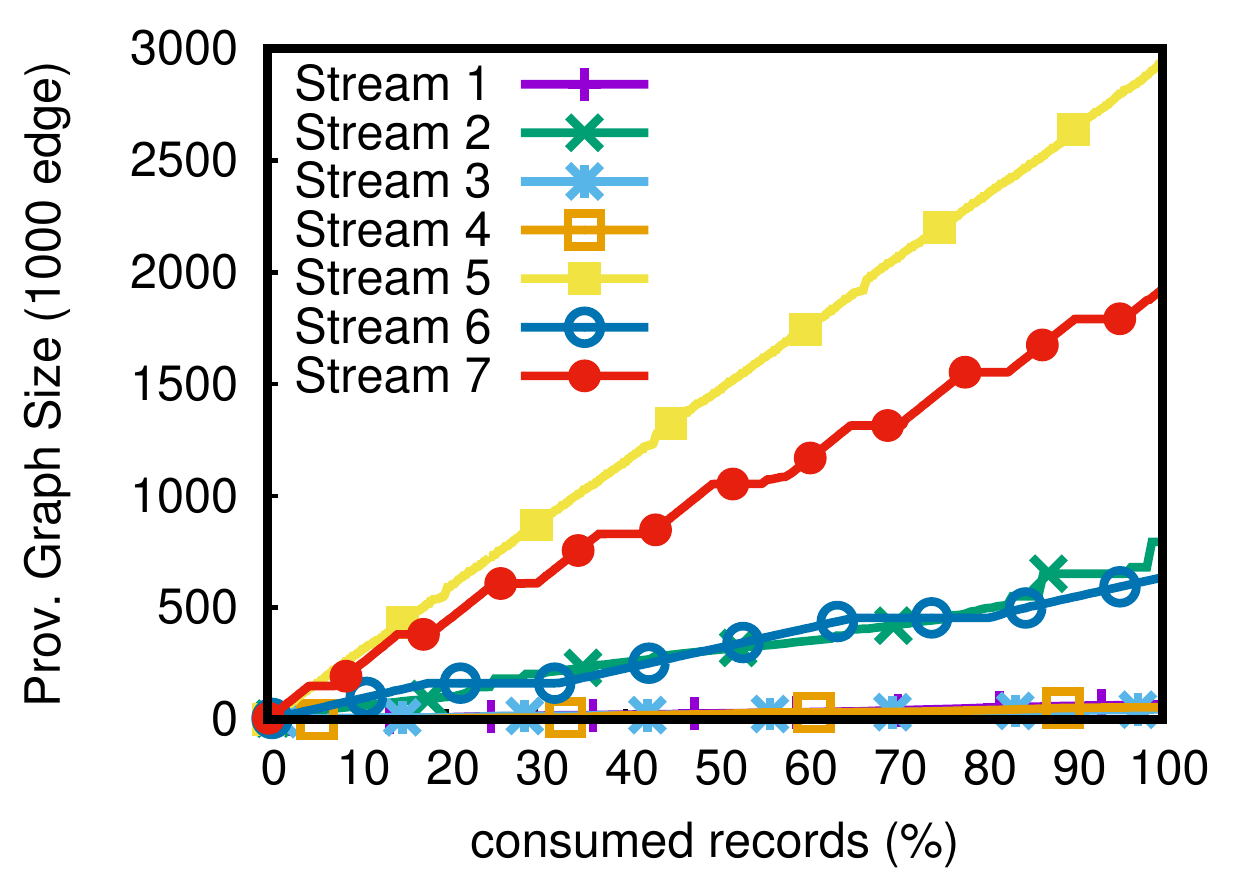}\label{fig:edges}
\hspace*{0.2\columnwidth}
    \includegraphics[width=.7\columnwidth]{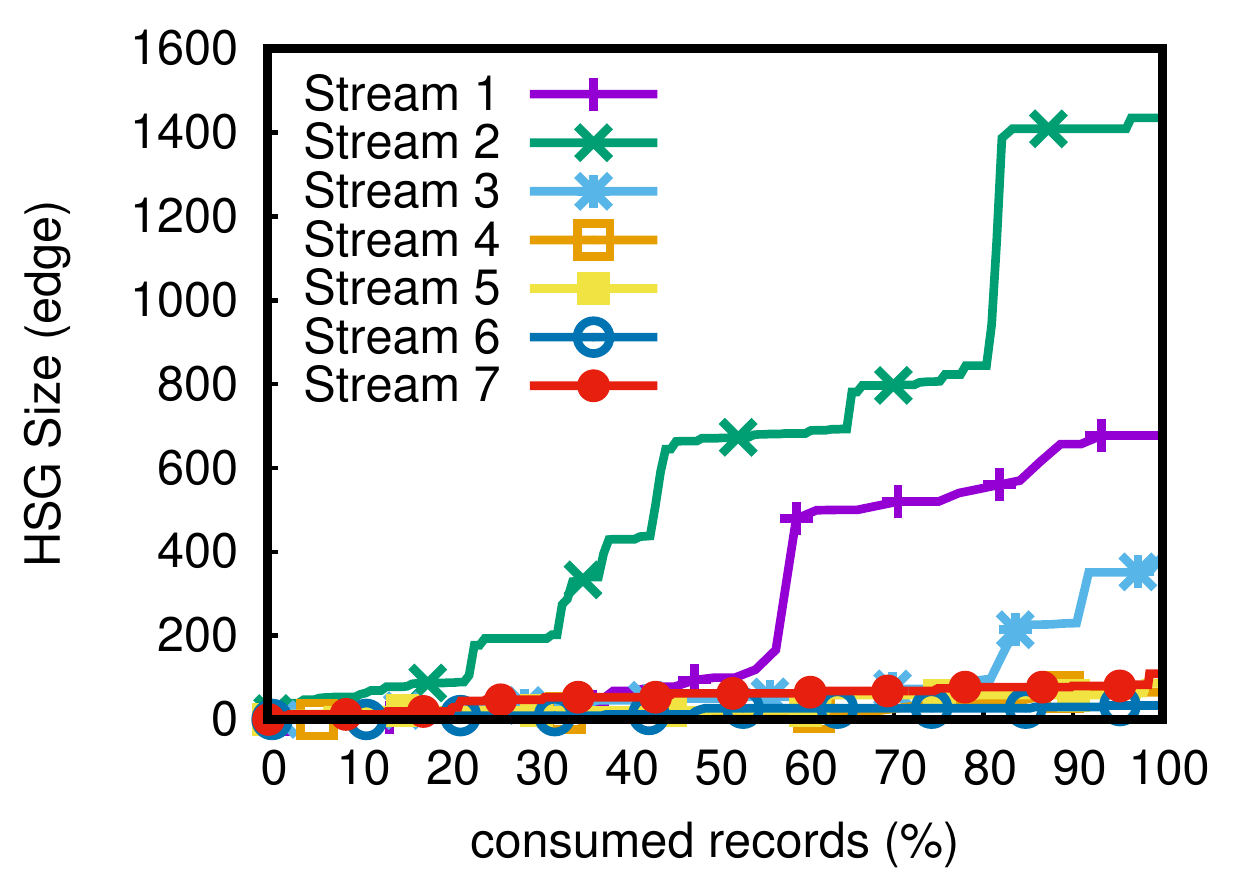}\label{fig:records}
  \end{center}
  \vspace*{-1em}
    \caption{(Left): Provenance graph growth vs. consumed records.  (Right): HSG growth vs. consumed records.}\label{fig:growth}
\end{figure*} 

\begin{figure*}[!ht]
  \begin{center}
    \includegraphics[width=.7\columnwidth]{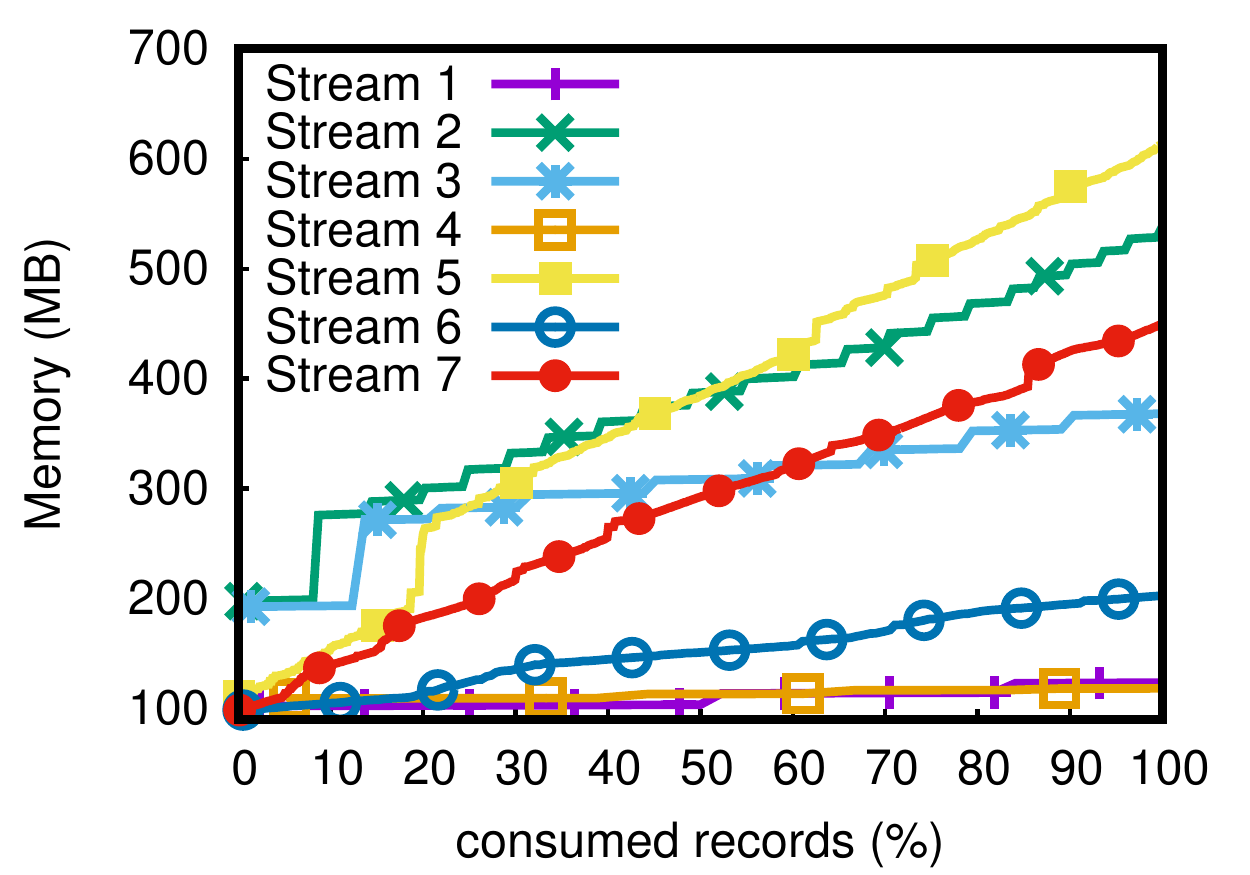}\label{fig:mem_per_record_scaled}
    \hspace*{0.2\columnwidth}
    \includegraphics[width=.7\columnwidth]{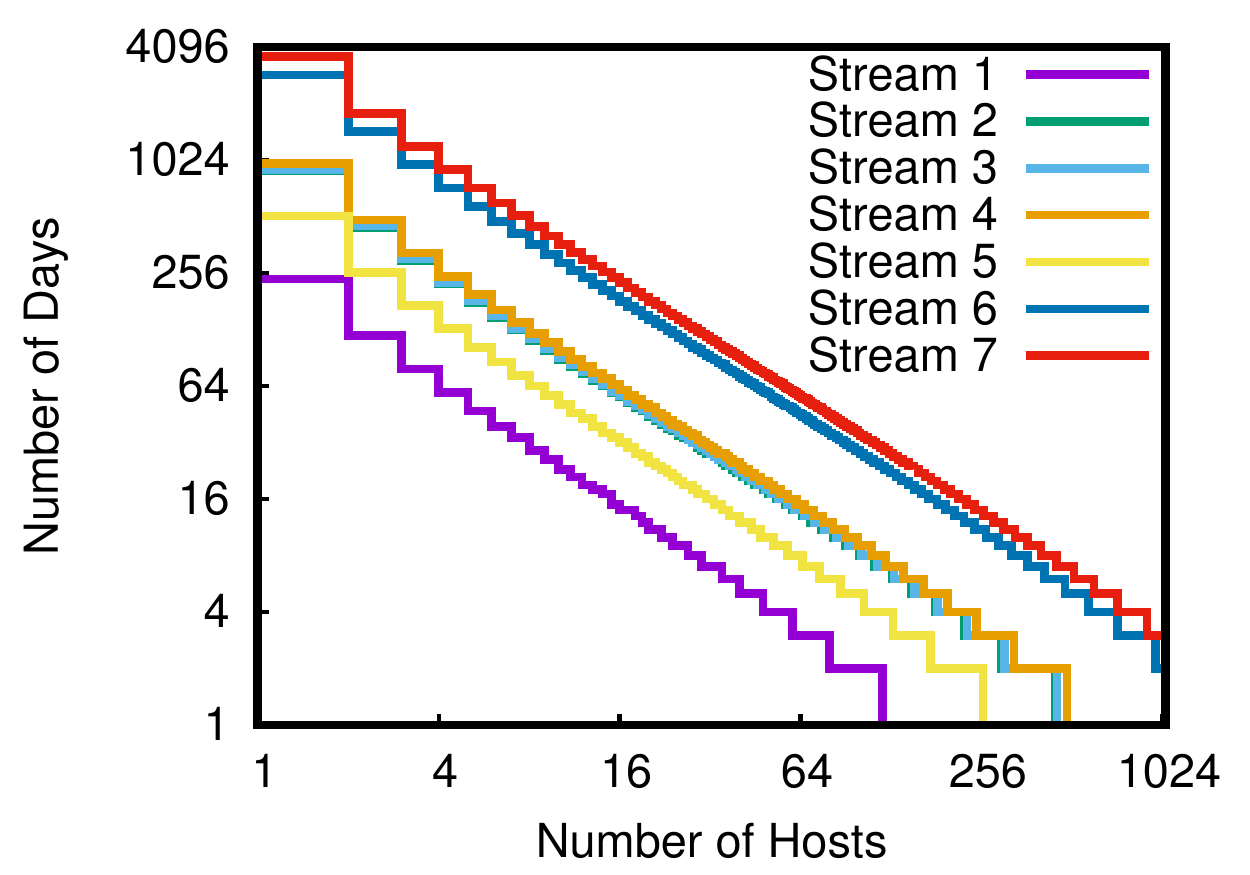}\label{fig:daily_mem}
  \end{center}
  \vspace*{-1em}
    \caption{(Left): Memory footprint (MB) vs. \% of records consumed.  (Right): Number of Days vs. extrapolated number of hosts that can be handled by \projname in respect to Memory consumption}\label{fig:mem}
    \vspace*{-1em}
\end{figure*}

\subsection{Attack Scenarios}\label{sec:eval:scenarios}
We now describe an additional attack scenario detected by \projname. For reasons of space, we include details of the rest of the scenarios and the related figures in the appendix. We note that Scenario-7.2 is discussed in section \ref{running} and a portion of its provenance graph and HSG are shown in Fig.s \ref{fig:provgraph} and \ref{fig:highlevelgraph}, respectively.

\noindent
{\bf Scenario-1: Drive-by Download.} In this attack scenario (see Fig. \ref{fig:drive-by-hsg}), the user visits a malicious website with a vulnerable Firefox browser. As a result, a file named {\em net} is dropped and executed on the victim's host. This file, after execution, connects to a C\&C server, and a reverse shell is provided to the attacker. The attacker then launches a shell prompt and executes commands such as {\em hostname}, {\em whoami}, {\em ifconfig}, {\em netstat}, and {\em uname}. Finally, the malicious executable exfiltrates information to the IP address of the C\&C server and then the attacker removes the dropped malicious file.

As can be seen from Fig. \ref{fig:drive-by-hsg}, in the Initial Compromise APT stage, an untrusted file is executed, which matches a TTP with the critical severity level.
The final {\em threat tuple} for this graph looks like $\langle  C,M,-,H,-,H,M \rangle$ for all APT stages (see Table \ref{results}). Consequently, the converted quantitative values are $\langle  10,6,1,8,1,8,6 \rangle$, which results in a {\em threat score} equal to 1163881. 

\begin{figure}[b]
  \begin{center}

    \includegraphics[width=.9\columnwidth,height=2in]{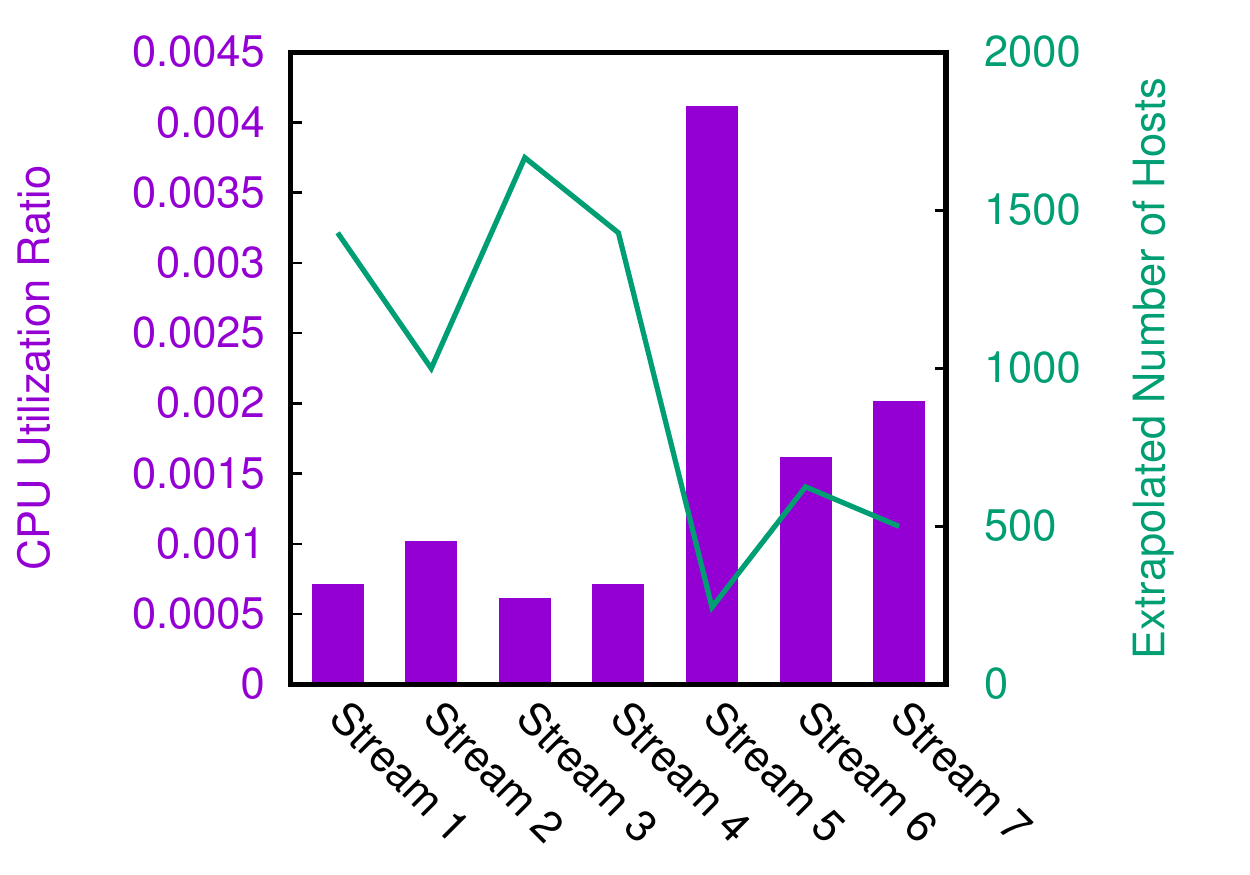}
  \end{center}
  \vspace*{-1em}
    \caption{CPU Utilization and the extrapolated number of hosts that can be handled by \projname in respect to CPU time.}
    \vspace*{-1em}
    \label{fig:cpu}
\end{figure}

\subsection{Finding the Optimal Threshold Value} \label{sec:eval-optimal-val}
To determine the optimal threshold value, we measured the precision and recall by varying threshold values as shown in Fig. \ref{fig:fscore}. F-score, the harmonic mean of precision and recall, is maximum at the interval [338.25, 608.26], which is the range from the maximum score of benign subgraphs to the minimum score of attack subgraphs. Therefore, by choosing any threshold in this range, \projname makes a clear distinction between attack and benign subgraphs in the tested datasets, with accuracy and recall equal to 1. 

To find the optimal value, we first transform the {\em threat scores} to a linear scale by getting their $n$th root, where $n$ equals to $\sum_{i=1}^{7} w_{i}$. The transformed value shows the average contribution of each APT step to the overall {\em threat score}, and it is a value in the range [1,10]. As all our tested datasets so far belong to single hosts, we exclude the weight of lateral movement step ($w_5$), which leads to $n= 8.3$. After getting the $n$th root, the interval of maximum F-score would change to [2.01, 2.16]. Finally, we consider the middle of this range ($2.09$) as the average severity that each APT step is allowed to contribute to the overall {\em threat score}, in a benign setting.

\subsection{Performance} \label{sec:eval:measure}
\noindent
{\bf Graph Size. } Fig. \ref{fig:growth} shows the comparison of the growth trends for provenance graph in thousands of edges (left) and the HSG in the number of edges (right). The graph size ratio measured in edges is 1875:1, i.e., an 1875-fold reduction is achieved in the process of mapping from the provenance graph to the HSG.

\noindent
{\bf Memory Use. } \projname was tested on an 8 core CPU with a 2.5GHz speed each
and a 150GB of RAM. Fig. \ref{fig:mem} (left) shows the memory consumption of
\projname with the number of audit records. It shows a nearly linear growth in
memory consumption since our system operates on audit records in-memory. Fig.
\ref{fig:mem} (right) shows extrapolation of how many hosts \projname can
support (regarding memory consumption) with scalability to an enterprise of
hundreds of hosts. It is evident that as the number of hosts is increased, the
duration that we can keep the full provenance graph in memory decreases. Notice
that both x and y-axes are in log-2 scale.

\noindent
{\bf Runtime. } While \projname consumes and analyzes audit records from a Kafka
server as the records become available in real-time, to stress-test its
performance, we assumed that all the audit records were available at once. Then,
we measured the CPU time for consuming the records, building the provenance
graph, constructing the HSG, and detecting APTs. We define ``CPU
Utilization'' as the ratio of required CPU time to the total duration of
a scenario. 
In Fig. \ref{fig:cpu}, the bars show CPU Utilization for each
scenario, and the line shows an extrapolation of how many hosts (of comparable
audit trace durations with the scenarios) \projname can support if CPU 
was the limiting factor. This chart shows that our single CPU can support an
enterprise with hundreds of hosts.

\subsection{Live Experiment} \label{sec:eval-live}
\begin{figure}[!ht]
  \begin{center}

    \includegraphics[width=.8\columnwidth]{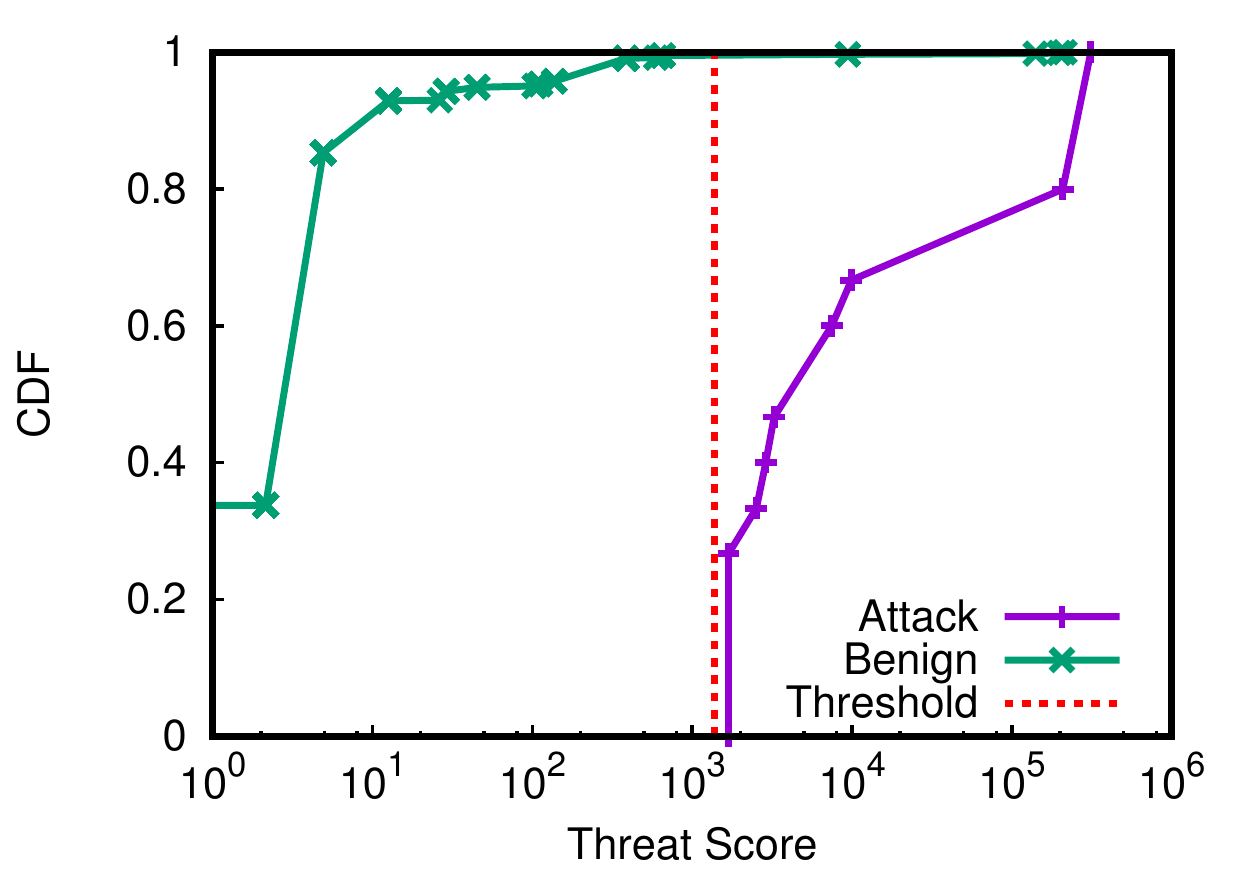}
  \end{center}
  \vspace*{-1em}
    \caption{Cumulative distribution function for attack vs. benign HSGs} 
    \vspace*{-1em}
    \label{fig:cdf}
\end{figure}

To explore how \projname would respond to attacks embedded within a  predominantly benign stream of events, we evaluated it as a live detection system. This experiment spanned 2 weeks, and during this period, audit logs of multiple systems, running Windows, Linux, or BSD, were collected and analyzed in real-time by \projname.
In this experiment, an enterprise is simulated with security-critical services such as  a web server, E-mail server, SSH server, and an SMB server for providing shared access to files. 
Similar to the previous datasets, an extensive set of normal activities are conducted during this experiment%
, and red-team carried out a series of attacks.
However, this time, we configured all the parameters beforehand and had no prior knowledge of the attacks planned by the red-team. Moreover, we had cross host internal connectivity, which makes APT stage 5 (Move\_laterally) a possible move for attackers. To this end, we set the detection threshold equal to $2.09^{\sum_{i=1}^{7} w_{i}}=2.09^{9.8}=1378$. Fig. \ref{fig:cdf} shows the cumulative distribution function for attack and benign HSGs that \projname constructs during this experiment. Note that there are some points representing {\em threat score} of benign HSGs, that have bypassed the threshold. We explain them as false positives in the following and then discuss some potential false negative scenarios.

\noindent
{\bf False Positives.} We noticed some false alarms because of SSH connections
made by system administrators. These connections come from untrusted IP
addresses, and subsequently, \projname aggregates the severity scores of all the
actions issued by the system administrator via an SSH connection. In some cases,
the {\em threat score} bypasses our threshold. The solution is to define a
custom tagging policy for servers such as ssh that perform authentication so
that the children of such servers aren't marked as untrusted
\cite{hossain2017sleuth}.

To further evaluate our system for false alarms, we also evaluated it on another
two weeks benign activity period. During this time, a diverse set of normal
activities were conducted, (including software updates and upgrades through
package managers) and \projname generated no false alarms.

Based on our results, we claim that the false positive  of \projname is at an acceptable rate considering the benefits it adds to an enterprise. Security analysts can manually check the raised alarms and neutralize HSGs that are falsely constructed.

\noindent
{\bf False Negatives.} Although we did not observe any false negatives during our experiments, here we discuss potential scenarios \projname might miss.

{\em Implicit causality between TTPs}: For information flow that avoids system calls, \projname have no direct visibility to the causal relations between system entities. However, if the rest of the attack unfolds with visibility through system calls, \projname will still partially reconstruct the attack.

{\em Multiple entry points}: As an active evasion technique, attackers might exploit multiple entry points that result in  detached subgraphs. \projname follows every single entry point until our detection threshold is satisfied and correlates TTPs from disjoint subgraphs when there is information flow between them. 
Nevertheless, some additional analyses might be needed to completely correlate attack steps, which are coming from different entry points and have no information flow in between.

%% file: relwork.tex
\section{Related Work}\label{sec:relatedwork}
\label{sec:relw}

\projname makes contributions to the problems of real-time alarm generation, alert correlation, and scenario reconstruction. A central idea in \projname is the construction and use of a high-level attack scenario graph as the underlying basis for all the above problems.    Below, we discuss related work in all of the above areas.

\noindent
{\bf Alarm Generation. } 
Host-based intrusion detection approaches fall under three classes: (1)
{\em misuse-based} \cite{kumar1995classification,porras1992penetration}, which detect behavior associated
with known attacks; (2) {\em anomaly-based} \cite{senseofself,datamining,sp01,execgraph,kruegelarg,manzoor2016fast,detectionWindows2015,xiaokuiccs}, which learn a model of benign behavior and detect deviations from it;
and (3) {\em specification-based} \cite{ko1997execution,raid01}, which detect attacks based on policies specified by experts. While the techniques of the first class cannot deal with unknown attacks, those of the second class can produce many false positives.
Specification-based techniques can reduce false positives, but they
require application-specific policies that are time-consuming to develop and/or
rely on expert knowledge. At a superficial level, the use of TTPs in \projname can be seen as an instance of misuse detection. However, our approach goes beyond classic misuse detection~\cite{kumar1995classification,porras1992penetration} in  the use of prerequisite-consequence patterns that are matched when there exist information flow dependencies between the entities involved in the matched TTP patterns. %

\noindent
{\bf Alarm Correlation.} 
Historically, IDSs have tended to produce alerts that are too numerous and
low-level for human operators. Techniques needed to be developed to
summarize these low-level alerts and greatly reduce their volume. 

Several approaches use alarm correlation to perform detection by clustering similar alarms and by identifying causal relationships between alarms
\cite{debar2001,ning2003,qin2003,noel2004,wang2008}. For instance, BotHunter \cite{bothunter} employs an anomaly-based approach to correlate dialog between internal and external hosts in a network.  {\sc Hercule} \cite{pei2016hercule} uses community
discovery techniques to correlate attack steps that may be dispersed across
multiple logs. Moreover, industry uses similar approaches for building SIEMs \cite{splunk,logrhythm,QRadar} for alert correlation and enforcement based on logs from disparate data sources. These approaches rely on logs generated by third-party applications running in user-space.  Moreover, alert correlation based on statistical features like alert timestamps does not help in precise detection of multi-stage APT attacks as they usually span a long duration. In contrast to these approaches, \projname builds on information flows that exist between various attack steps  for the purpose of alert correlation. The use of kernel audit data in this context was first pursued in ~\cite{zhai2006integrating}. However, differently from \projname, that work is purely misuse-based, and its focus is on using the correlation between events to detect steps of an attack that are missed by an IDS. \projname uses the same kernel audit data but pursues a different approach based on building a main-memory dependency graph with low memory footprint, followed by the derivation of an HSG based on the high-level specification of TTPs to raise alerts, and finally correlate alerts based on the information flow between them. An additional line of work on alert correlation relies on the proximity of alerts in time~\cite{kruegel2004intrusion}. \projname, in contrast, relies on information flow and causality connections to correlate alerts and is therefore capable of detecting even attacks where the steps are executed very slowly.

\noindent
{\bf Scenario Reconstruction.} A large number of research efforts have been 
focused on generation and use of system-call level logs in forensic 
analysis, investigation and recovery \cite{bates2015trustworthy,gehani2012spade,taser2005,forensix,pohly2012hi,king2003backtracking,king2005enriching,liu2018towards,wang2018fear,lee2013high,ma2016protracer,ma2017mpi,sadegh2018propatrol}. Most forensic analysis approaches trace back from a given compromise event to determine the causes of that compromise. Among these, BEEP \cite{lee2013high}, ProTracer \cite{ma2016protracer}, and MPI \cite{ma2017mpi} use training and code instrumentation and annotations to divide process executions into smaller units, to address dependency explosion and provide better forensic analysis. PrioTracker \cite{liu2018towards} performs timely causality analysis by quantifying the notion of event rareness to prioritize the investigation of abnormal causal dependencies.
In contrast, \projname uses system event traces to perform \emph{real-time detection}, with  integrated forensics capabilities in the detection framework, in the form of high-level attack steps, without requiring instrumentation. 

Recent studies \cite{pasquierruntime,Shu:2018:TIC:3243734.3243829,hossain2017sleuth} have used system-call level logs for real-time analytics. 
{\sc Sleuth} \cite{hossain2017sleuth} presents tag-based techniques for attack detection and in-situ
forensics. \projname makes several significant advances over {\sc Sleuth}.
First, it shows how to address the dependence explosion problem by using the
concept of minimum ancestral cover and developing an efficient algorithm for
its incremental computation. Second, {\sc Sleuth}'s scenario graphs are at the
same level of abstraction as the provenance graph, which can be too low-level
for many analysts, and moreover, lacks the kind of actionable information in
HSGs. Third, {\sc Sleuth}'s graphs can become too large on long-running attacks,
whereas \projname generates compact HSGs by using noise reduction and
prioritization techniques.

\noindent
{\bf Attack Granularity. } Sometimes, the coarse granularity of audit logs may limit reasoning about information flows. For example, if a process with a previously loaded sensitive file is compromised, the attacker can search for sensitive content inside its memory region without using system calls. However, when such information is exfiltrated, \projname correlates the exfiltration with the other actions of that process (i.e., the sensitive file read) and eventually raises an exception. Furthermore, \projname can be adapted to take advantage of additional works, which 
track information flows at finer granularities, 
either by instrumenting additional instructions \cite{angelos,flowdroid} or by decoupling taint tracking \cite{kwon18mci,ming2016straighttaint,chow2008decoupling,ji2017rain}. Such fine-grained information flow tracking can provide much more precise provenance information at the cost of performance overheads.

%% file: conclusion.tex
\section{Conclusion} \label{sec:conclusion}

We present \projname, a real-time APT detection system that correlates tactics, techniques, and procedures that might be used to carry out each APT stage. 
\projname generates a high-level graph that summarizes the attacker's steps in real-time. We evaluate \projname against nine real-world APT threats 
 and deploy it as a real-time intrusion detection tool. The results show 
 that \projname successfully detects APT campaigns with high precision and low false alarm rates.

%% file: appendix.tex
\section*{Appendix}

\begin{figure}[h!]
  \begin{center}

    \includegraphics[width=\columnwidth]{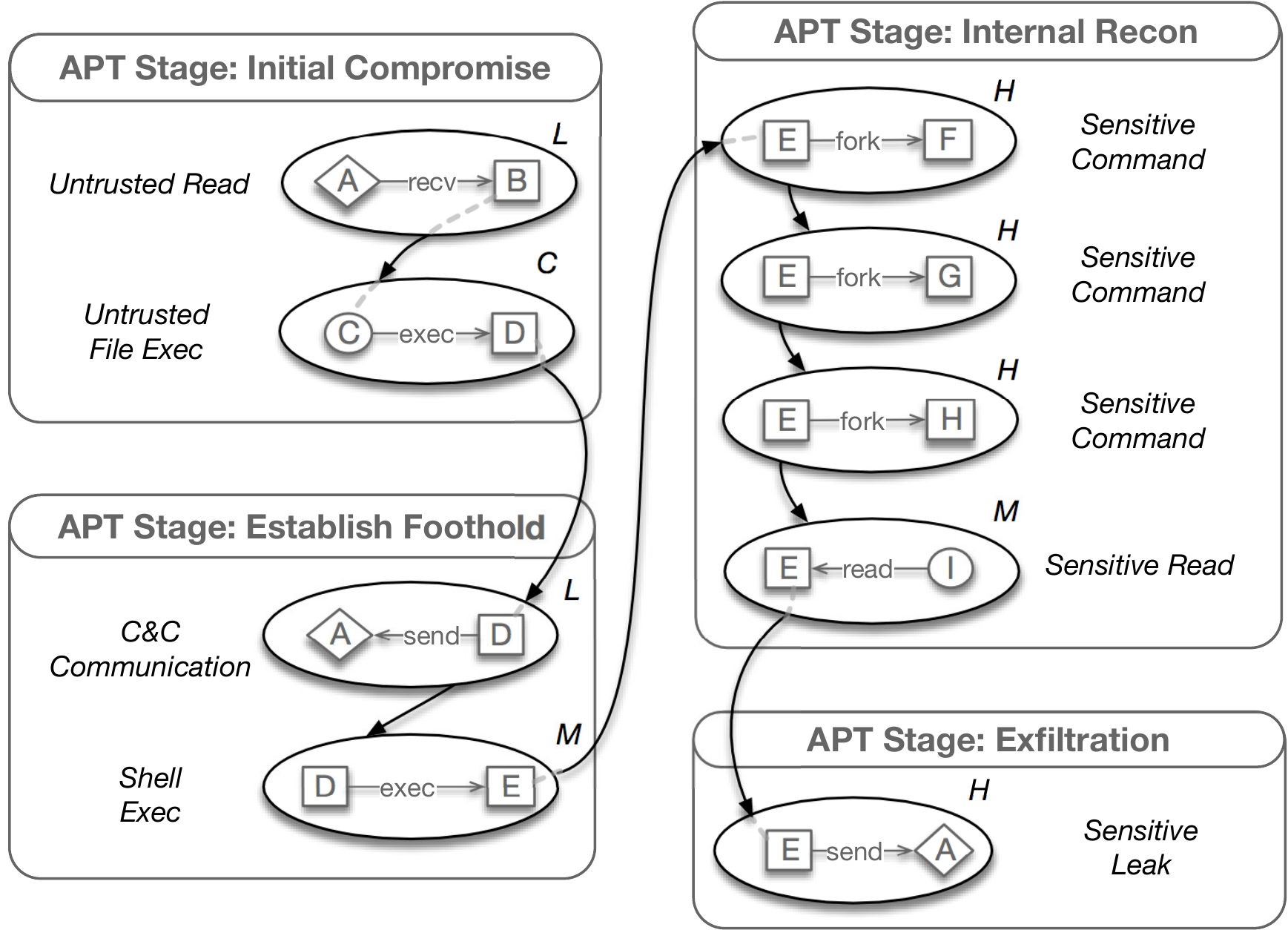}
  \end{center}
  \vspace*{-1em}
    \caption{HSG of Scenario-2. Notations: A= Untrusted External Address; B= Firefox; C= Trojan File (diff); D= Executed Trojan Process; E= /bin/dash; F= ifconfig; G= hostname; H= netstat; I= password.txt;}
    \label{fig:scenario_2_HSG}
\end{figure}

{\bf Scenario-2: Trojan.} This attack scenario (Fig. \ref{fig:scenario_2_HSG}) begins with a user downloading a malicious file. The user then executes the file. The execution results in a C\&C communication channel with the attacker's machine. The attacker then launches a shell  and executes some information gathering commands such as {\em hostname}, {\em whoami}, {\em ifconfig}, {\em netstat}, and {\em uname}. Finally, the attacker exfiltrates some secret files. Note that this attack scenario is similar to the {\em Drive-by Download} scenario discussed earlier except that the initial compromise happens via a program that the user downloads. Another important insight from the detection results of this scenario is that it was missing important events that are relevant to the C\&C communication ({\em connect}) and final cleanup ({\em unlink}) activity of the attack. Even with such incomplete data, \projname was able to flag this as an APT since the Threat score surpassed the threshold.

\begin{figure}[h!]
  \begin{center}

    \includegraphics[width=\columnwidth]{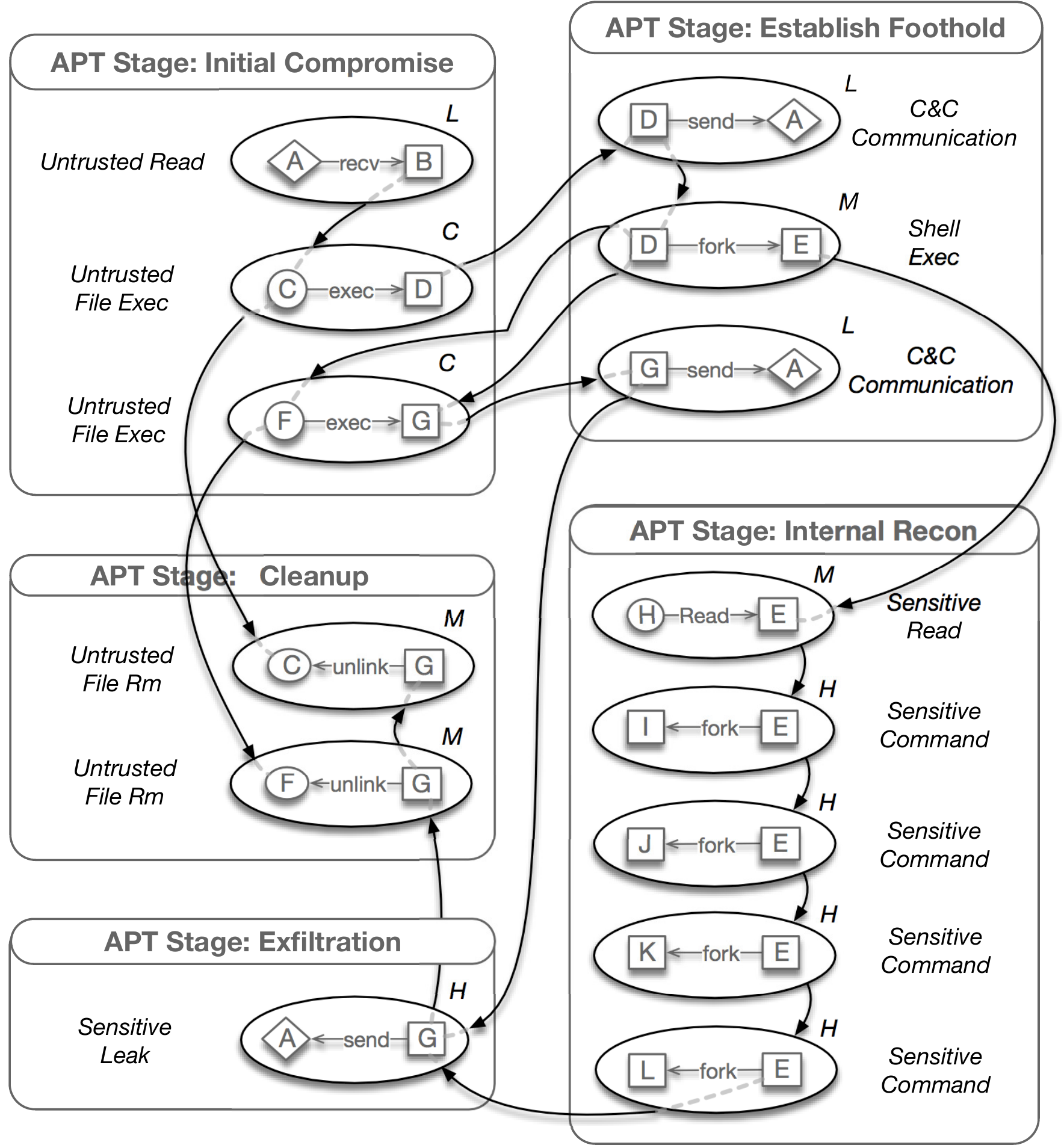}
  \end{center}
  \vspace*{-1em}
    \caption{HSG of Scenario-3. Notations: A= Untrusted External Address; B= Firefox; C= Trojan File (tedit); D= Executed Trojan Process; E= /bin/dash; F= Malicious Executable file (py); G= Executed Malicious Process; H= password.txt; I= whoami; J= ifconfig; K= netstat; L= uname;}
    \label{fig:scenario_3_HSG}
\end{figure}

{\bf Scenario-3: Trojan.} In this attack (Fig. \ref{fig:scenario_3_HSG}), a user is convinced to download a malicious Trojan program (texteditor) via Firefox. Next, the user moves the executable file to another directory, changes its name (tedit), and finally executes it. After the execution, a C\&C channel is created, and a reverse shell is provided to the attacker. The attacker launches a shell prompt and executes information gathering commands like {\em hostname},  {\em whoami}, {\em ifconfig}, and {\em netstat}. The attacker then deploys another malicious file, exfiltrates information, and finally cleans up his footprints. This scenario differs from {\em Trojan-1} because it has an additional activity that remotely deploys a new malicious executable.

\begin{figure}[h!]
  \begin{center}

    \includegraphics[width=\columnwidth]{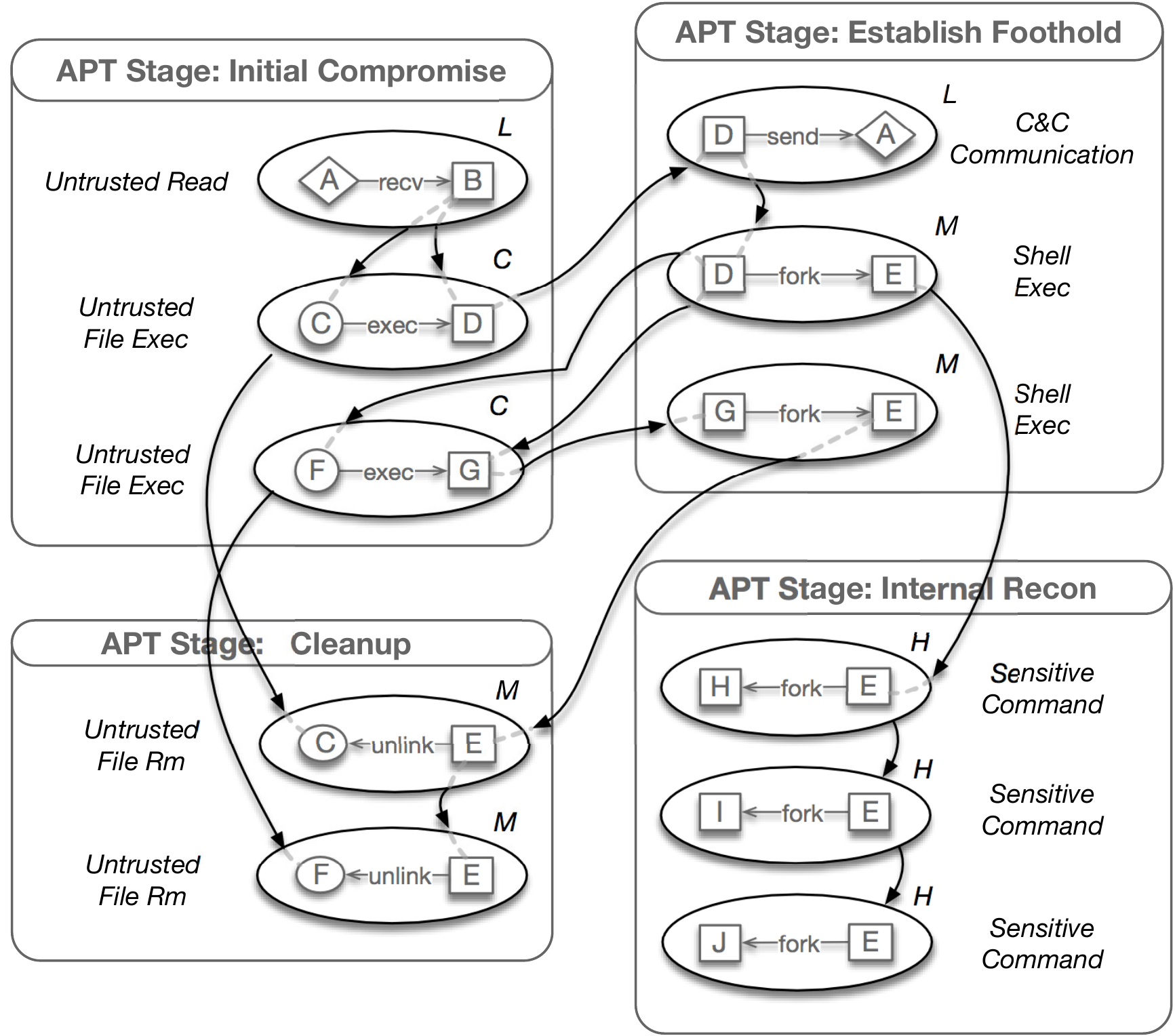}
  \end{center}
  \vspace*{-1em}
    \caption{HSG of Scenario-4. Notations: A= Untrusted External Address; B= Firefox.exe; C= Malicious dropped file (procman.exe); D= Executed Malware Process; E= cmd.exe; F= Malicious Batch file (burnout.bat); G= Executed Batch Process; H= hostname; I= whoami; J= ipconfig;}
    \label{fig:scenario_4_HSG}
\end{figure}

{\bf Scenario-4: Spyware.} This attack (Fig. \ref{fig:scenario_4_HSG}) begins when the red-team compromises Firefox. The user on the victim host then loaded a hijacked remote URL. Next, a shellcode from the URL is executed to connect to a C\&C server from which it downloaded a malicious binary, wrote it to disk, and executed it. The execution of the malicious binary results in a reverse shell channel for C\&C communications. The attacker then ran the shell command, resulting in a new {\em cmd.exe} process and a new connection to the C\&C server. The operator ran reconnaissance commands ({\em hostname}, {\em whoami}, {\em ipconfig}, {\em netstat}, {\em uname}). The attacker then exfiltrated the {\em password.txt} file and then deleted it. Finally, the malicious binary drops a batch file that deletes attack footprints, including the malicious binary itself.

\begin{figure}[h!]
  \begin{center}

    \includegraphics[width=\columnwidth]{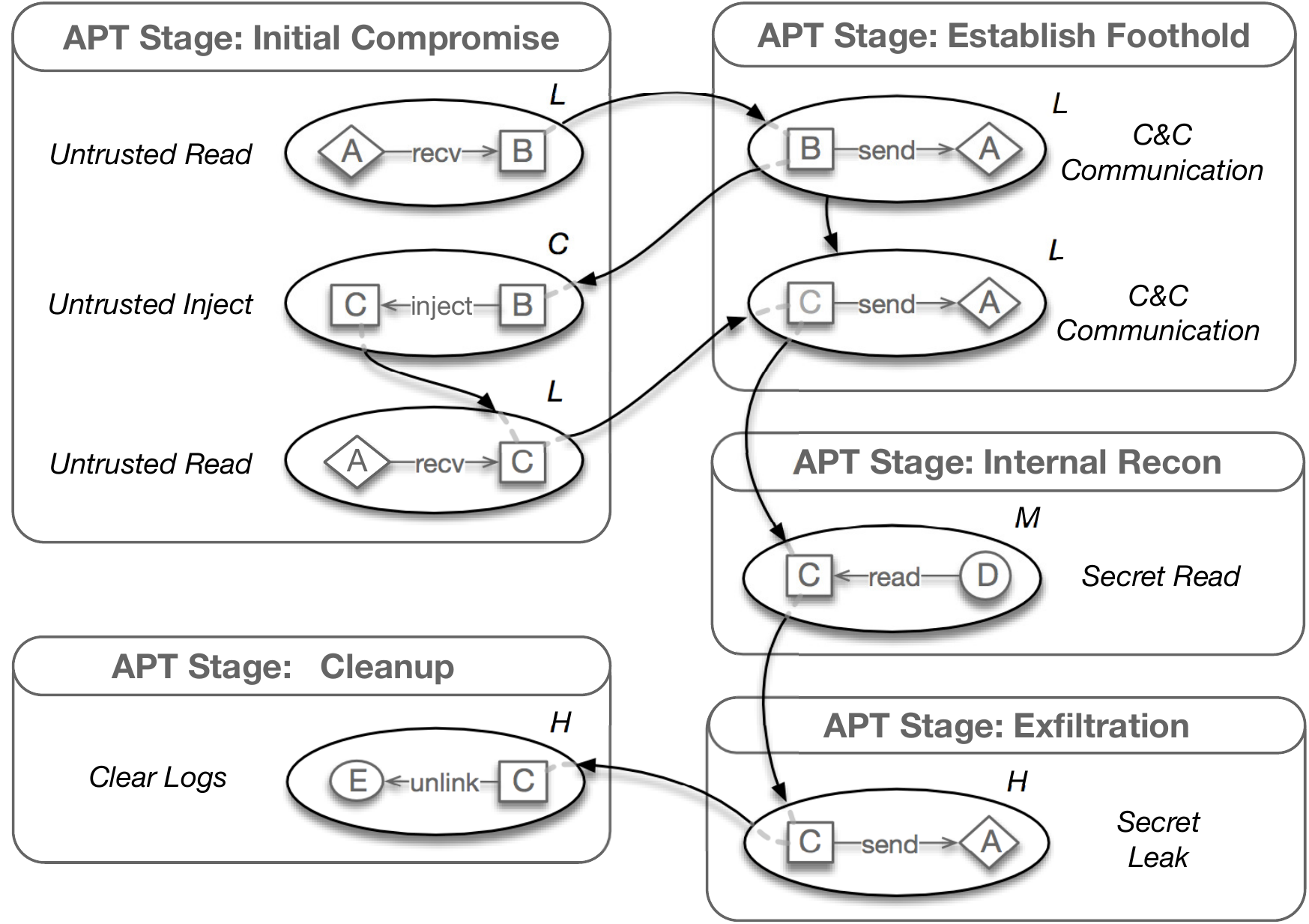}
  \end{center}
  \vspace*{-1em}
    \caption{HSG of Scenario-5.1 (Eternal Blue). Notations: A= Untrusted External Address; B= lsass.exe; C= rundll32.exe; D= password.txt; E= Winevt logs;}
    \label{fig:eternal_blue}
\end{figure}

{\bf Scenario-5.1: Eternal Blue. }
This APT exploits vulnerable SMB \cite{smb} services in Windows. In this scenario (see Fig. \ref{fig:eternal_blue}), Meterpreter \cite{meterpreter} was used with the recently implemented Eternal Blue exploit and Double Pulsar reflective loading capabilities. The attacker exploited the listening SMB service on port 445 of the target. A shellcode was then downloaded and executed on the target. The shellcode performed process injection into the {\em lsass.exe} process. {\em lsass.exe} then launched {\em rundll32.exe}, which connected to the C\&C server and downloaded-and-executed Meterpreter. Next, Meterpreter exfiltrated a sensitive file and cleared Windows event logs.

\begin{figure}[h!]
  \begin{center}

    \includegraphics[width=\columnwidth]{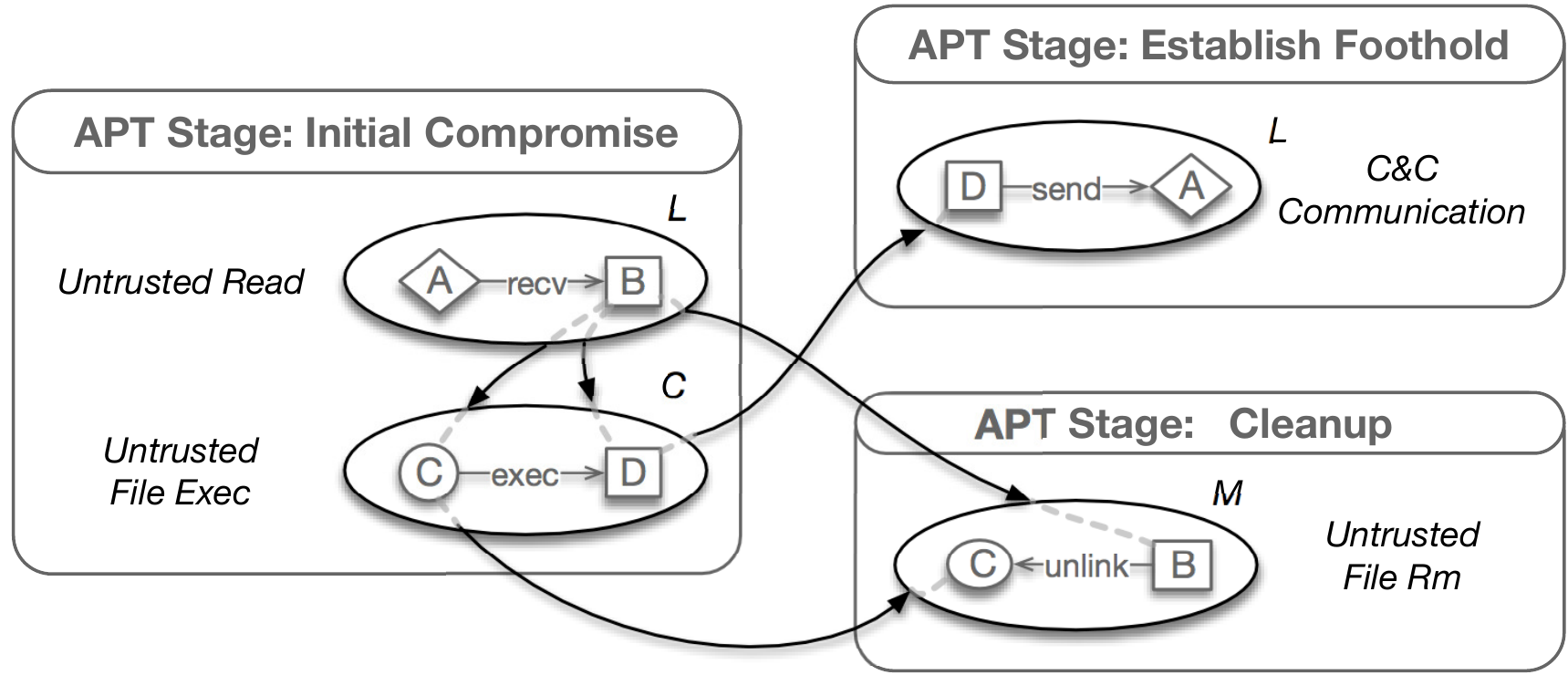}
  \end{center}
  \vspace*{-1em}
    \caption{HSG of Scenario-5.2. Notations: A= Untrusted External Address; B= Firefox.exe; C= Malicious dropped file (spd.exe); D= Executed Malware Process;}
    \label{fig:scenario_6_HSG}
\end{figure}

{\bf Scenario-5.2: RAT.}
In this attack (Fig. \ref{fig:scenario_6_HSG}), Firefox navigates to a malicious website and gets exploited.
Then, a Remote Access Trojan (RAT) is uploaded to the victim's machine and executed. After execution, a connection to the C\&C server has happened, and the malicious RAT is deleted. This attack scenario is incomplete, and no harm is done.

\begin{figure}[h!]
  \begin{center}

    \includegraphics[width=\columnwidth]{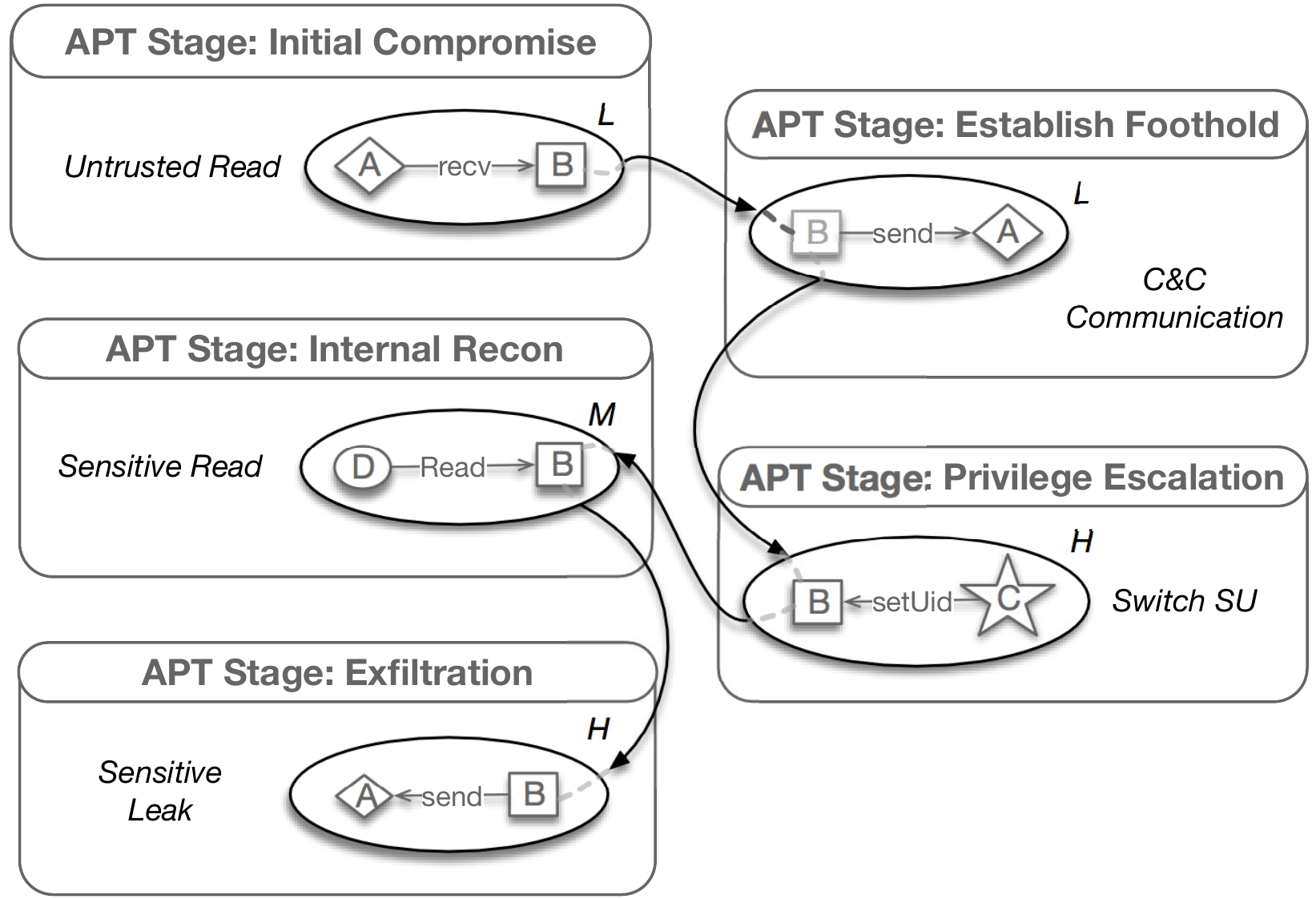}
  \end{center}
  \vspace*{-1em}
    \caption{HSG of Scenario-6. Notations: A= Untrusted External Address; B= Nginx; C= Root userID; D= Passwd.txt;}
    \label{fig:scenario_7_HSG}
\end{figure}

{\bf Scenario-6: Web-Shell.} The assumption in this attack (Fig. \ref{fig:scenario_7_HSG}) is that {\em Nginx} web server has a vulnerability that gives the attacker access to run arbitrary commands on the server (similar to Shellshock bug). As a result, the attacker exfiltrates a sensitive file.
The important insight here is that  by capturing sufficiently strong APT signals of an ongoing attack through TTP matching, \projname accurately flags an APT, even when a critical APT step is missing (initial compromise in this case).

\begin{figure}[h!]
  \begin{center}

    \includegraphics[width=\columnwidth]{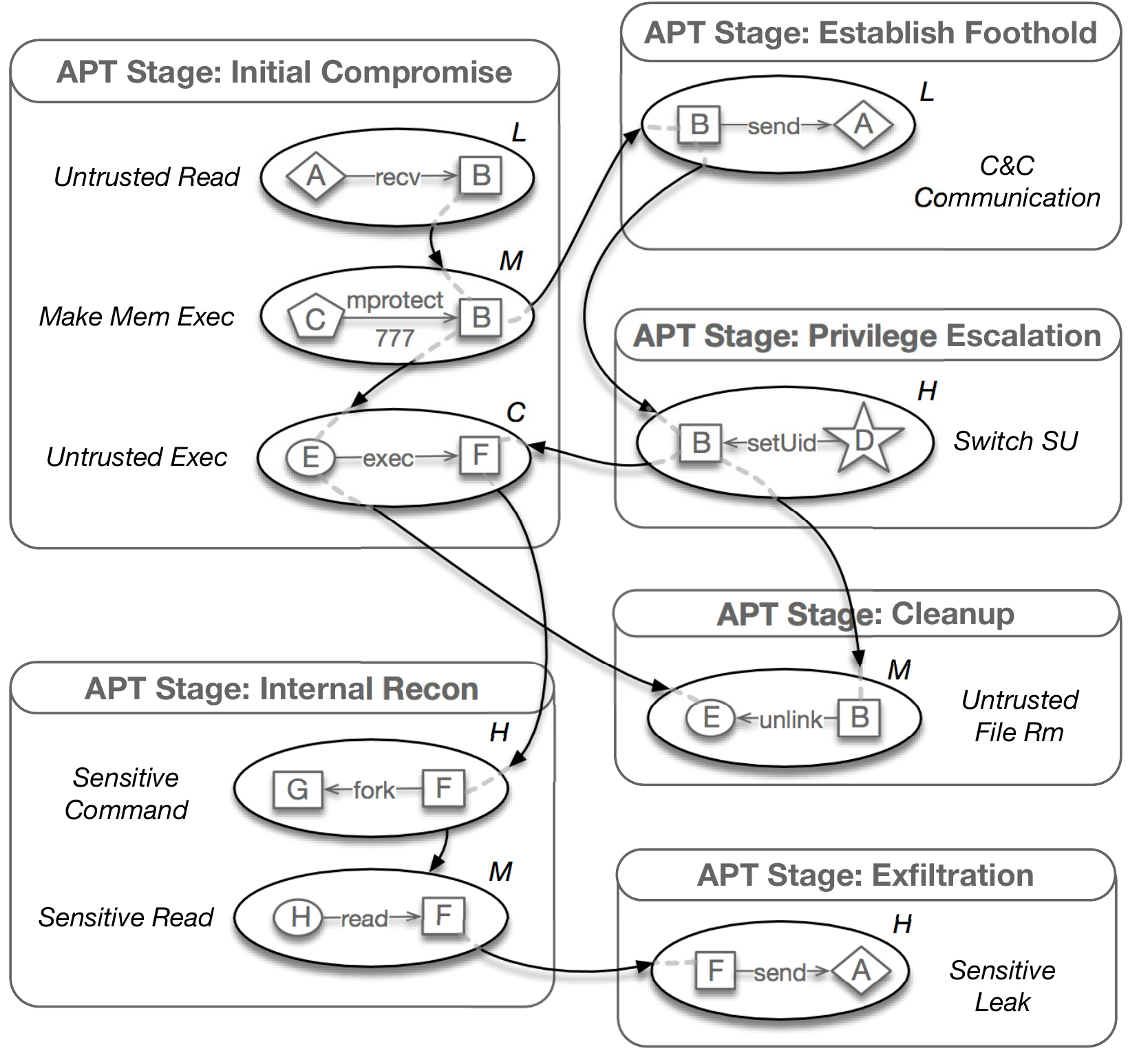}
  \end{center}
  \vspace*{-1em}
    \caption{HSG of Scenario-7.1. Notations: A= Untrusted External Address; B= Nginx; C= Memory; D= Root userID; E= Malicious dropped file (py); F= Executed Malware Process; G= uname; H= /etc/shadow;}
    \label{fig:scenario_8_HSG}
\end{figure}

{\bf Scenario-7.1: RAT.} A vulnerable {\em Nginx} server was installed during the setup period. The attacker exploits the {\em Nginx} server by throwing a malicious shell-code. {\em Nginx} runs the malicious shell-code which results in the download and execution of a malicious RAT. Next, RAT connects to a C\&C server and gives administrative privileges to the remote attacker.  The attacker remotely executes some commands. It then deploys some malicious Python scripts and exfiltrates information. The HSG of this attack is shown in Fig. \ref{fig:scenario_8_HSG}.

